\documentclass[journal]{IEEEtran}
\usepackage{subfig}
\usepackage{booktabs}
\usepackage{cite}
\usepackage{makecell}
\usepackage{amsmath,amssymb,amsfonts,amsthm}
\usepackage{algorithmic}
\usepackage{graphicx}
\usepackage{stfloats} 
\usepackage{multirow}
\usepackage[hang,flushmargin]{footmisc}
\usepackage{textcomp}
\usepackage{xcolor}
\usepackage{tikz}
\usetikzlibrary{shapes,arrows}
\usepackage[ruled,vlined]{algorithm2e}
\usepackage{booktabs}

\newtheorem{theorem}{Theorem}
\newtheorem{proposition}{Proposition}
\newtheorem{lemma}{Lemma}

\def\BibTeX{{\rm B\kern-.05em{\sc i\kern-.025em b}\kern-.08em
    T\kern-.1667em\lower.7ex\hbox{E}\kern-.125emX}}
\begin{document}

\title{Broad and Spectral-Efficient Beamforming for the Uni-polarized Reconfigurable Intelligent Surfaces\\
 
}
\author{Mohammad Javad-Kalbasi, Student Member, IEEE, Mohammed Saif, Member, IEEE, and Shahrokh Valaee,  Fellow, IEEE

	\thanks{Mohammad Javad-Kalbasi, Mohammed Saif, and Shahrokh Valaee are with the Department of Electrical and Computer Engineering, University of Toronto, Toronto, Canada, 
Email: mohammad.javadkalbasi@mail.utoronto.ca, mohammed.saif@utoronto.ca,  valaee@ece.utoronto.ca.

This work was supported in part by funding from the Innovation for Defence Excellence and Security (IDEaS) program from the Department of National Defence (DND).		
	}
	
	\vspace{-.35cm}
 }
 
\maketitle
\IEEEpubidadjcol

\begin{abstract}
A reconfigurable intelligent surface (RIS) is composed of low-cost elements that manipulate the propagation environment from a transmitter by intelligently applying phase shifts to incoming signals before they are reflected. This paper explores a uni-polarized RIS with linear shape aimed at transmitting a common signal to multiple user equipments (UEs) spread across a wide angular region. To achieve uniform coverage, the uni-polarized RIS is designed to emit a broad and spectral-efficient beam featuring a spatially flat-like array factor, diverging from the conventional narrow beam approach. To achieve this objective, we start by deriving probabilistic lower and upper bounds for the average spectral efficiency (SE) delivered to the UEs. Leveraging the insights from the lower bound, we focus on optimizing the minimum value of the power domain array factor (PDAF) across a range of azimuth angles from \(-\frac{\pi}{2}\) to \(\frac{\pi}{2}\). We employ the continuous genetic algorithm (CGA) for this optimization task, aiming to improve the SE delivered to the UEs while also creating a wide beam. Extensive simulation experiments are carried out to assess the performance of the proposed code, focusing on key metrics such as the minimum and average values of the PDAF and the SE delivered to the UEs. Our findings demonstrate that the proposed code enhances the minimum SE delivered to the UEs while maintaining the desired attribute of a broad beam. This performance is notably superior to that of established codes, including the Barker, Frank, and Chu codes. 
\end{abstract}


\section{Introduction}
A reconfigurable intelligent surface (RIS) is a key enabling technology for a smarter radio environment \cite{RIS_Mohanad, Javad_globecom2023, saifglobecom}, which can control the propagation environment by smartly tuning phase shifts to direct impinging signals in a desired way. Since its envision, RIS has been widely studied in the literature to improve different metrices in various ways, e.g., energy efficiency \cite{Javad_globecom2023}, user equipment (UE) localization \cite{M}, network connectivity \cite{saifglobecom_E}, and coverage \cite{RIS_Mohanad, 9293155}.

In the context of beamforming, the deployment of RIS plays a crucial role in achieving both broad and spectral-efficient communication. Traditional beamforming techniques often face limitations in coverage and efficiency due to the fixed nature of conventional antenna arrays. However, the dynamic reconfigurability of RIS offers a transformative solution. Specifically, by intelligently adjusting the phase shifts of its elements, RIS can form highly directional beams that enhance signal strength and reduce interference, thus improving spectral efficiency (SE). Moreover, RIS can create multiple simultaneous beams, broadening coverage and serving multiple UEs efficiently. This dual capability of enhancing SE and broadening coverage makes RIS an indispensable technology for next-generation wireless networks, promising significant improvements in data rates, energy consumption, and overall network performance. 

\subsection{Motivation and Related Works}
The previous works on RIS-assisted communications consider enhancing the performance of individual UEs by configuring phase shifts of a RIS to form narrow beams towards them \cite{9293155, saifglobecom_E}. Such narrow beams are  effective at targeting specific areas or directions with high precision. It has been observed that there are many practical applications where multiple UEs, located in a wide angular area and possibly at unknown locations, tend to receive a common signal, necessitating an efficient broad beams approach. Some cases of such practical situations where broad beams approach is needed are: (i) when a sender transmits cell-specific signals to receivers, such as primary synchronization signal, secondary synchronization signal, or cell-specific reference signal via RIS \cite{ramezani}; (ii) UEs have a common interest in downloading popular files, especially videos \cite{saif}; (iii) small-packet transmission, low-latency transmission, and high mobility scenarios \cite{highmob}. For instance, autonomous vehicles operate in highly dynamic environments that require continuous and reliable communication for safety and efficient navigation. %\ac{Remove:Traditional communication systems often struggle to adapt quickly enough to such rapidly changing conditions and may not provide necessary coverage.} 
This wide coverage and dynamic system requires an efficient broad beams approach, which is essential for detecting and interacting with other vehicles, roadside infrastructure, and non-vehicular obstacles. The implementation of this broad beamforming can significantly enhance vehicular safety. For example, in situations where an autonomous vehicle must rapidly communicate emergency braking due to sudden obstacles, the RIS-enabled system can broadcast this critical information across a broad beam, covering all directions. This ensures that all nearby vehicles, regardless of their position relative to the sender, receive the warning promptly and can react accordingly. Such reliable and efficient communication capabilities also facilitate better traffic management, allowing vehicles to share information about traffic conditions and optimal routes in real-time. For such practical situations, one might use a beam sweeping approach along with a RIS forming narrow beams towards possible angular directions. This approach is inefficient in terms of resource utilization, since it needs to continuously sweep across the area and sequentially directing beams towards different angular directions. Consequently, it requires rapid reconfiguration of the RIS to ensure coverage across all desired angles. In addition, each change in direction might require recalibration and realignment, consuming significant time and computational resources.

In the literature, several methods for broadening the beam reflected by the RIS have been proposed in different RIS-assisted systems, e.g., \cite{partialcover, widen, ramezani, plannarRIS}. In particular, the study by \cite{partialcover} aims to achieve quasi-static broad coverage by minimizing the difference between a predefined pattern and the power pattern reflected by the RIS. Concurrently, \cite{widen} explores the creation of a wide beam codebook while imposing constraints on the sidelobe levels. Despite their contributions, these proposed methods fall short in generating a perfectly broad beam, even within a confined angular range. As will be elaborated upon later in this paper, a perfectly broad beam is characterized by a power domain array factor (PDAF) that remains constant across the entire range of interest.

Achieving an efficient flat beam using conventional RIS that deploys uni-polarized beamforming is not possible, unless only a single element is activated \cite{ramezani}. Recently, a dual-polarized beamforming is studied in \cite{Dual1, Dual2} to exploit the polarization degree of dual-polarized array antennas. The main idea is to design beamforming weights of dual-polarized antennas such that different polarization beams complement each other, yielding an overall broad radiated beam. The work in \cite{ramezani} utilizes linear dual-polarized RIS to propose a phase shift design that enables flat beamforming from RIS. It is shown that the phase shift vectors of RIS elements in horizontal and vertical polarizations should form a Golay complementary sequence pairs \cite{Golaypair} for generating a flat beam. The authors extend the results to the case of plannar dual-polarized RISs in \cite{plannarRIS}, and show that a perfectly uniform radiated beam is only achievable if the auto-correlation function (ACF) of the RIS configuration matrices add up to a Kronecker delta function, a property possessed by the so-called Golay complementary array pairs \cite{Golaypair, planRIS}. Although dual-polarized RISs offer the flexibility to control and manipulate both horizontal and vertical polarizations, which can be crucial for applications requiring polarization diversity to enhance channel capacity or to mitigate polarization-dependent fading and interference, the choice between uni-polarized and dual-polarized RIS should be guided by the specific requirements of the deployment scenario and the desired balance between complexity, cost, and performance. Thus, one may be tempted to use a uni-polarized RIS  for broad beamforming instead of a dual-polarized RIS for the following reasons.
\begin{itemize}
\item \textbf{Simplicity in Design and Fabrication:} Uni-polarized RISs are generally simpler to design and manufacture compared to dual-polarized RISs. This simplicity arises from the need to manage only one polarization state, which reduces the complexity of the elements and the control mechanisms required to manipulate the incident electromagnetic waves.

\item \textbf{Cost Effectiveness:} Since they are less complex, uni-polarized RISs tend to be more cost-effective in terms of both production and maintenance. The reduced complexity in the electronic and structural design translates into lower production costs and potentially higher reliability due to fewer components that could fail.

\item \textbf{Efficiency in Specific Applications:} For applications where only one polarization is relevant or required, a uni-polarized RIS can be more efficient. This is because it focuses all its resources on optimizing the manipulation of electromagnetic waves in one polarization state, potentially achieving better performance in that state than a dual-polarized RIS, which must split its capabilities between two polarizations \cite{uniadvan}.

\item \textbf{Reduced Interference:} In scenarios where interference is a concern, controlling only one polarization can help in minimizing the interference effects. A uni-polarized RIS can be designed to be less susceptible to cross-polarization interference, which can be advantageous in densely populated frequency environments \cite{crossinter}.
\end{itemize}
Motivated by the above-mentioned main advantages of using a uni-polarized RIS over a dual-polarized RIS, this paper is interested in  broad beamforming for the uni-polarized RIS.
\subsection{Contribution}
Building on the advantages previously discussed, this paper introduces a novel code for configuring phase shifts in a uni-polarized, linear-array RIS to generate a wide and spectrally efficient beamform. Utilizing continuous genetic algorithm (CGA), this code optimizes the minimum value of the PDAF across all azimuth angles from \(-\frac{\pi}{2}\) to \(\frac{\pi}{2}\). Furthermore, we examine both the average PDAF and the SE delivered to the UEs. To highlight the importance of optimizing PDAF for achieving high SE levels, we derive probabilistic lower and upper bounds for the average SE, which are based on both the minimum and the average PDAF values.

We benchmark our proposed code against established codes, such as the Barker code \cite{barkerref}, the Frank code \cite{francref}, and the Chu code \cite{chucodd}. These codes are known for their delta-like ACFs and are ideal for generating broad beams. To demonstrate the efficacy of our proposed code, we conduct Markov-Chain Monte-Carlo (MCMC) simulations to assess the minimum and average values of SE delivered to the UEs. The results show that our proposed code outperforms traditional codes in terms of minimum PDAF and minimum SE while maintaining satisfactory average values. This significant improvement in the minimum values highlights the robustness and reliability of the proposed code, ensuring better performance under worst-case conditions and proving its viability for applications requiring broad and spectrally efficient beamforms.

Next, in Section \ref{S}, the system model is presented. Then flat beam condition and well-known codes are reviewed in Section \ref{Flat}. In Section \ref{SE}, we establish the probabilistic lower and upper bounds on the average SE. In Section \ref{PS}, the proposed code is formally derived and important metrics for evaluating the performance of a broadbeam are introduced. The numerical results are given in Section \ref{NR}. Finally, Section \ref{C} concludes the paper.

\section{System Model}\label{S}
In this paper, we consider a downlink RIS-assisted communication system with one transmitter and a set $\mathcal K=~\{1, 2, \ldots, K\}$ of $K$ UEs. The transmitter intends to transmit a common signal to the UEs through a uni-polarized RIS with linear shape. Assuming a dense urban scenario, where direct links between the transmitter and the UEs are blocked, and communication can only occur through the RIS, RIS can establish reliable communication with the blocked UEs to deliver this common message over broad beamforming.\footnote{Although we do not consider direct links between the transmitter and the UEs, the proposed framework still holds for the scenario where direct links exist. With a minor adaptation, our proposed code can then be incorporated with the direct links, demonstrating the robustness and applicability of our code.} Similar to \cite{ramezani}, we assume that the transmitter and the UEs are equipped with one antenna each, while the RIS has $M$ elements with a horizontal uniform linear array (ULA) topology.  In an RIS, each meta-atom consists of a reflector that reflects signals, and can independently modify the phase of the impinging wireless signals.
%Fig. \ref{DUAL} illustrates the structure of a ULA RIS.

Let $\mathbf h \in \mathbb C^M$ and $\mathbf g \in \mathbb C^M$ represent the transmitter-RIS and RIS-UE channels, respectively. We consider a simple yet reasonably accurate line-of-sight (LoS) channel between the transmitter and the RIS. The LoS channels can be expressed as $\mathbf h=~\sqrt{\beta_h G_0(\theta_{h})} \mathbf a(\theta_{h})$ and $\mathbf g= \sqrt{\beta_g G_0(\theta)} \mathbf a {(\theta)}$, where $\beta_h$ and $\beta_g$ denote the
path-loss at the reference RIS,  $\theta_{h}$ denotes the angle of arrival (AoA) from the transmitter to the RIS, which is assumed to be known and fixed at the deployment stage, $\theta$ is the angle of departure (AoD) from the RIS to a typical UE, $G_0(\cdot)$ is the radiation power pattern of a single RIS element, and
$\mathbf a(\cdot)$ is the RIS array response vector. Further, we denote $\mathbf{\Phi}=[\phi_{1},\ldots, \phi_{m}, \ldots, \phi_{M}]$ as the phase shift configuration vector of the RIS, where $\phi_{m}$ denotes the phase shift applied by the $m$-th RIS element to the
incident signal. Let us define the vector $\mathbf{\Psi}=[\psi_{1},\ldots,\psi_{m},\ldots,\psi_{M}]$, where $\psi_{m}=e^{j\phi_{m}}$. Thus,  the received signal at an arbitrary
UE can be written as\footnote{We assume a LoS channel between the transmitter and the RIS, which is the
most practical scenario \cite{wireapp}. We intentionally consider LoS between the RIS and the UEs because we want to achieve flat-like array gain in all LoS directions.} 
\begin{equation} \label{eq1}
y=\sqrt {P} \mathbf g^T \text{diag}\left(\mathbf{\Psi}\right) \mathbf h s+ n,
\end{equation}
where $\text{diag}\left(\mathbf{\Psi}\right)$ is a diagonal matrix with diagonal elements $\mathbf{\Psi}$, $s \backsim \mathcal N_{\mathbb C}(0,1)$ is the transmitted signal, $P$ is the transmit power, and $n \backsim \mathcal N_{\mathbb C}(0, \sigma^2)$ is the receiver noise, where $\sigma^2$ denotes the noise power. The array response vector can be expressed as \cite{arrayresp}
\begin{equation*}
\mathbf{a}(x)=[1,e^{-j\frac{2\pi \Delta}{\lambda} \sin{x}}, \ldots,  e^{-j\frac{(M-1)2\pi \Delta}{\lambda} \sin{x}}]^T,
\end{equation*}
where $\Delta$ is the inter-element spacing between adjacent RIS elements and $\lambda$ is the wavelength of the transmitted signal. We can alternatively rewrite (\ref{eq1}) as
\begin{equation*}
y=\sqrt{P G_0(\theta_{h}) G_0(\theta)\beta_h \beta_g} \mathbf{\Psi} (\mathbf{a}(\theta_{h})\odot\mathbf{a}(\theta))s+n,
\end{equation*}
where the notation \(\mathbf{a}(\theta_{h}) \odot \mathbf{a}(\theta)\) represents the element-wise product between the vectors \(\mathbf{a}(\theta_{h})\) and \(\mathbf{a}(\theta)\). Then, the signal-to-noise ratio (SNR) at a typical UE can be written as
\begin{equation*}
\text{SNR}=\frac{P \beta_h \beta_g}{\sigma^2} G_0(\theta_{h}) G_0(\theta) A(\mathbf{\Phi},\theta),
\end{equation*}
where 
\begin{eqnarray}\label{arrayHV}
A(\mathbf{\Phi},\theta)&=& \left| \mathbf{\Psi}(\mathbf{a}(\theta_{h})\odot \mathbf{a}(\theta))\right|^2 \\ \nonumber 
~&=& \left|\sum\limits_{m=1}^{M} e^{j\phi_{m}}e^{-j\frac{2\pi \Delta (m-1)}{\lambda}\left(\sin{\theta_{h}}+\sin{\theta}\right)} \right|^2
\end{eqnarray}
represents the PDAF of the RIS. The total radiation power pattern observed by a typical UE is given by $G(\mathbf{\Phi},\theta)=~G_0(\theta_{h}) G_0(\theta)A(\mathbf{\Phi},\theta) $.

Our aim in this paper is to design a coding scheme that ensures the RIS radiated signal forms a broad and spectrally efficient beam, covering the entire angular range where potential UEs may be located. Before presenting our proposed code in Section \ref{PS}, we will first derive a condition necessary for generating a flat-like beam and then review several well-known coding schemes in the next section.
 
\section{Flat-like beamforming}\label{Flat}
First, we derive a condition for radiating a flat-like beam from the RIS. Let us define
\begin{equation}\label{trans}
\alpha=\frac{2\pi \Delta}{\lambda}\left(\sin{\theta_{h}}+\sin{\theta}\right).
\end{equation}
%and
%\[\varphi_{m}=e^{j\phi_{m}}, ~~~~~\text{for}~~ m=1,2,\ldots,M.\]
Then, the PDAF can be written as
\begin{equation}\label{conditiono}
A(\mathbf{\Phi},\alpha)={\left|\sum_{m=1}^{M}e^{j\phi_{m}}e^{-j(m-1)\alpha}\right|}^{2}.
\end{equation}
The term represented in (\ref{conditiono}) corresponds to the squared (spatial) discrete-time Fourier transforms of the sequence $\mathbf{\Psi}$, when this sequence is zero-padded. Given that the power spectral density (PSD) of a sequence is equal to the square of its Fourier transform \cite{fourierpair}, (\ref{conditiono}) can be rephrased as
\begin{equation*}
A(\mathbf{\Phi},\alpha)=S_{\mathbf{\Psi}}(\alpha),
\end{equation*}
where $S_{\mathbf{\Psi}}(\alpha)$ denotes the PSD of $\mathbf{\Psi}$. Noting Wiener-Khinchin theorem, the PSD and ACF form Fourier transform pairs \cite{fourierpair}. Thus, we have
\begin{equation*}
F\{R_{\mathbf{\Psi}}[\tau]\}=S_{\mathbf{\Psi}}(\alpha),
\end{equation*}
where $R_{\mathbf{\Psi}}[\tau]$ indicates the ACF of the vector $\mathbf{\Psi} \in \mathbb{C}^{M}$, which is given by
\begin{equation}\label{autfun}
R_{\mathbf{\Psi}}[\tau] =
\begin{cases}
    \sum\limits_{m=1}^{M-\tau}\psi_{m}\psi_{m+\tau}^{\star}, & \text{if}~ \tau= 0,\ldots,M-1 \\\\
    \sum\limits_{m=1}^{M+\tau}\psi_{m-\tau}\psi_{m}^{\star}, & \text{if}~ \tau=-M+1,\ldots,-1\\\\
    0, & \text{otherwise.}
\end{cases}
\end{equation}
It is straightforward to see that the peak value of the ACF occurs at zero lag, expressed as 
\[R_{\mathbf{\Psi}}[0] = \sum\limits_{m=1}^{M} |\psi_{m}|^{2} = \sum\limits_{m=1}^{M} |e^{j\phi_{m}}|^{2}=M.\] 

Moreover, noting (\ref{autfun}), we conclude that 
\[R_{\mathbf{\Psi}}[M-1]=R^{\star}_{\mathbf{\Psi}}[-M+1]=\psi_{1}\psi_{M}^{\star}=e^{j(\phi_{1}-\phi_{M})} \neq 0.\]
This indicates that achieving zero for all sidelobes of the ACF is unfeasible. Consequently, the ACF cannot perfectly mimic a delta function, leading to the conclusion that the PDAF cannot be uniformly flat. Although it is not possible to create a perfectly flat beam, we can produce a beam that approximates flatness by reducing the sidelobe levels in the ACF, thus achieving flat-like beams. To accomplish this, given that the PDAF and ACF are Fourier transform pairs, it is essential that \( R_{\mathbf{\Psi}}[\tau] \) behaves like a delta function with respect to \(\tau\), implying minimal sidelobes in the function. Notable codes that meet this criterion include the Barker, Frank, and Chu codes \cite{barkerref}, \cite{francref}, \cite{chucodd}. Next, we briefly review the structure of these well-known codes.

\subsection{Barker Code}
In telecommunications, the Barker code refers to a finite string of digital values possessing the optimal autocorrelation characteristic. It serves as a synchronization pattern, facilitating alignment between the sender and the receiver in a bitstream. It is established that the absolute value of sidelobes in the Barker code is limited to at most $1$ \cite{barkerref}, rendering it a favorable choice for generating a broad PDAF. There are only nine known Barker codes, with each having a maximum length of $M$ up to $13$. These known Barker codes are given in Table \ref{Barkcod}. The sidelobe level ratio is calculated as \( 20\log_{10}\left(\frac{1}{M}\right) \) in dB.
\begin{table}[t]
\centering
\caption{Known Barker Codes}
\label{Barkcod}
\begin{tabular}{@{}lll@{}}
\toprule
Length ($M$)  & Codes ($\mathbf{\Phi}^{\text{Barker}}$)& Sidelobe Level Ratio \\ \midrule
$2$ & $(0,\pi)$ , $(0,0)$   & $-6$ dB    \\
$3$ & $(0,0,\pi)$    & $-9.5$ dB     \\
$4$ & $(0,0,\pi,0)$, $(0,0,0,\pi)$     & $-12$ dB     \\
$5$ & $(0,0,0,\pi,0)$     & $-14$ dB     \\
$7$ & $(0,0,0,\pi,\pi,0,\pi)$     & $-16.9$ dB    \\
$11$ & $(0,0,0,\pi,\pi,\pi,0,\pi,\pi,0,\pi)$     & $-20.8$ dB    \\
$13$ & $(0,0,0,0,0,\pi,\pi,0,0,\pi,0,\pi,0)$     & $-22.3$ dB    \\ \bottomrule
\end{tabular}
\end{table}

\subsection{Frank Code}
Another family of codes that has delta-like ACF is called the Frank code \cite{francref}, which represents a polyphase code modulation format utilized for pulse compression. It employs harmonically related phases derived from specific fundamental phase increments. Let us define the square $N \times N$ matrix as
\[\Omega=\left[\frac{2\pi}{N}(i-1)(j-1)~ (\text{mod} ~2\pi)\right],\]
for  $i=1,2,\ldots,N$ and $j=1,2,\ldots,N$. Then the Frank code with length $M=N^{2}$ is given by \cite{francref}
\begin{equation}\label{Francformul}
\mathbf{\Phi}^{\text{Frank}}=\left(\Omega^{(1)},\Omega^{(2)},\ldots,\Omega^{(N)}\right),
\end{equation}
where $\Omega^{(i)}$ is the $i$-th row of matrix $\Omega$ for  $i=1,2,\ldots,N$.
For instance, in the case of the Frank code with $N = 4$, the matrix $\Omega$ is
\[
\Omega=\begin{bmatrix}
    0 & 0 & 0 & 0\\
    0 & \frac{\pi}{2} & \pi & \frac{3\pi}{2} \\
    0 & \pi & 0 & \pi \\
    0 & \frac{3\pi}{2} & \pi & \frac{\pi}{2}
\end{bmatrix}
\]
Thus, the Frank code with length $16$ is expressed by the sequence
\[\mathbf{\Phi}^{\text{Frank}}=\left(0,0,0,0,0,\frac{\pi}{2},\pi,\frac{3\pi}{2},0,\pi,0,\pi,0,\frac{3\pi}{2},\pi, \frac{\pi}{2}\right).\]
\subsection{Chu Code}
Frank codes are constrained by the requirement that their lengths should be perfect squares. In contrast, Chu codes \cite{chucodd}, renowned for their beneficial autocorrelation characteristics, do not have this limitation on code lengths. The definition of this family of codes is as follows
\begin{equation}\label{autfunn}
\phi_{m}^{\text{Chu}} =
\begin{cases}
    \frac{q\pi {(m-1)}^{2}}{M}~ (\text{mod} ~2\pi), & \text{if}~ M~\text{is even,} \\\\
    \frac{q\pi m(m-1)}{M}~ (\text{mod} ~2\pi), &  \text{if}~ M~\text{is odd},
\end{cases}
\end{equation}
where $m=1,2,\ldots,M$ and $q$ is a positive integer relatively prime to $M$. The Chu code is essential to the 3GPP Long Term Evolution (LTE) air interface, being employed in several key components, including the Primary Synchronization Signal (PSS), the random access preamble (PRACH), the uplink control channel (PUCCH), the uplink traffic channel (PUSCH), and the sounding reference signals (SRS). The Chu code is employed to design the phase shift of RIS elements, aiming to produce a PDAF with a flat-like appearance.
 
\begin{figure}[t]
    \centering
    %\vspace{0.5cm} % Add vertical space between rows
    \begin{minipage}[b]{0.5\columnwidth}
        \centering
        \includegraphics[width=0.9\linewidth]{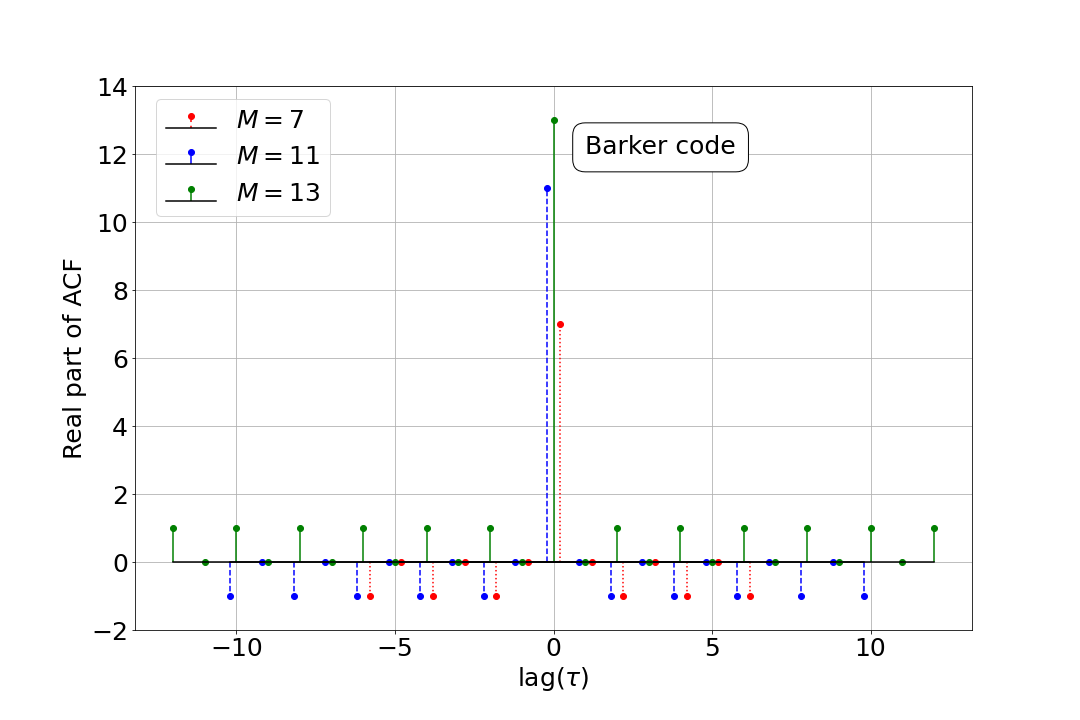}
        %\caption*{Barker code}
        %\label{fig:subfig1}
    \end{minipage}\hfill
    \begin{minipage}[b]{0.5\columnwidth}
        \centering
        \includegraphics[width=0.9\linewidth]{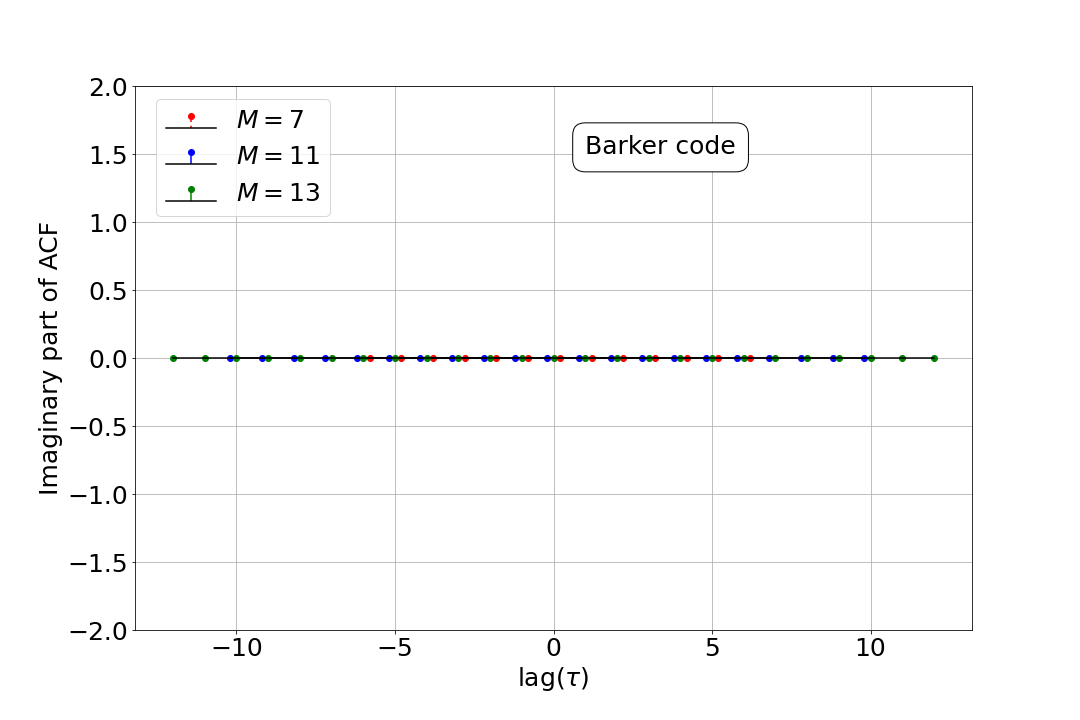}
        %\caption*{Barker code}
        %\label{fig:subfig2}
    \end{minipage}
    
    %\vspace{0.5cm} % Add vertical space between rows
    
    \begin{minipage}[b]{0.5\columnwidth}
        \centering
        \includegraphics[width=0.9\linewidth]{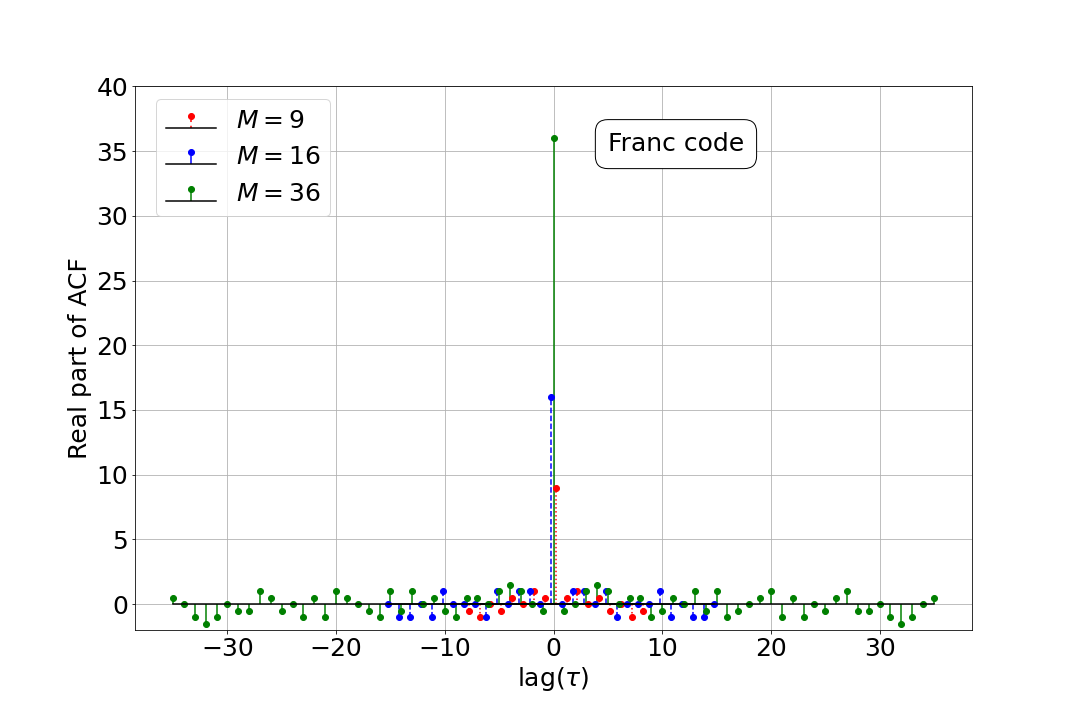}
        %\caption*{Franc code}
        %\label{fig:subfig3}
    \end{minipage}\hfill
    \begin{minipage}[b]{0.5\columnwidth}
        \centering
        \includegraphics[width=0.9\linewidth]{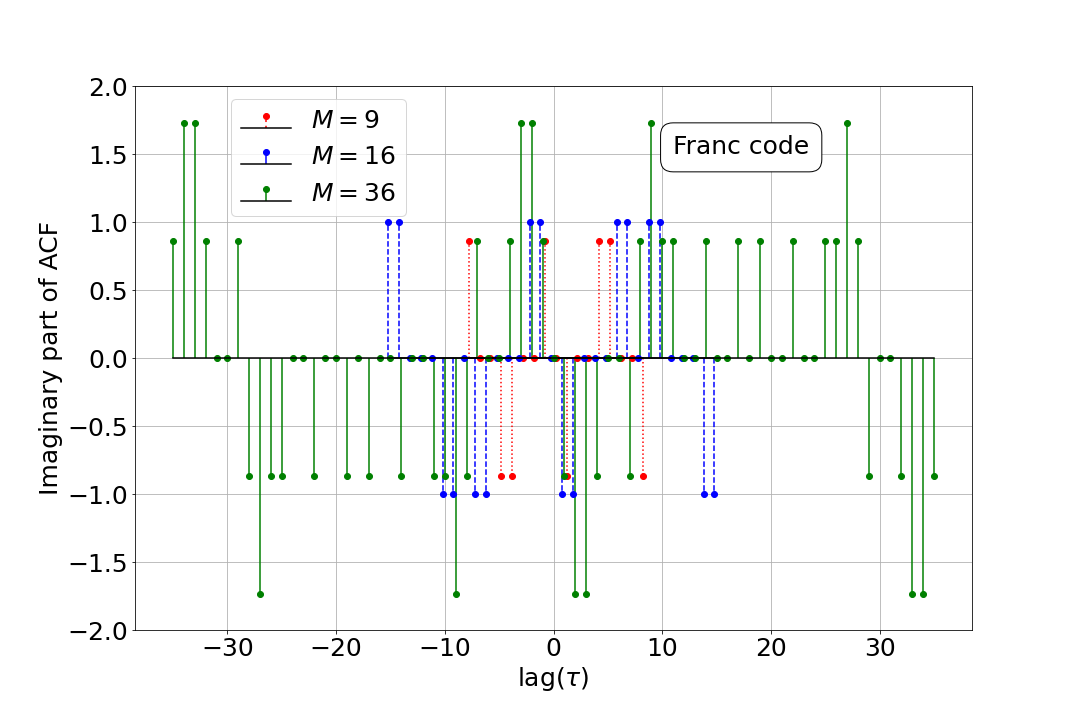}
        %\caption*{Franc code}
        %\label{fig:subfig4}
    \end{minipage}
     \begin{minipage}[b]{0.5\columnwidth}
        \centering
        \includegraphics[width=0.9\linewidth]{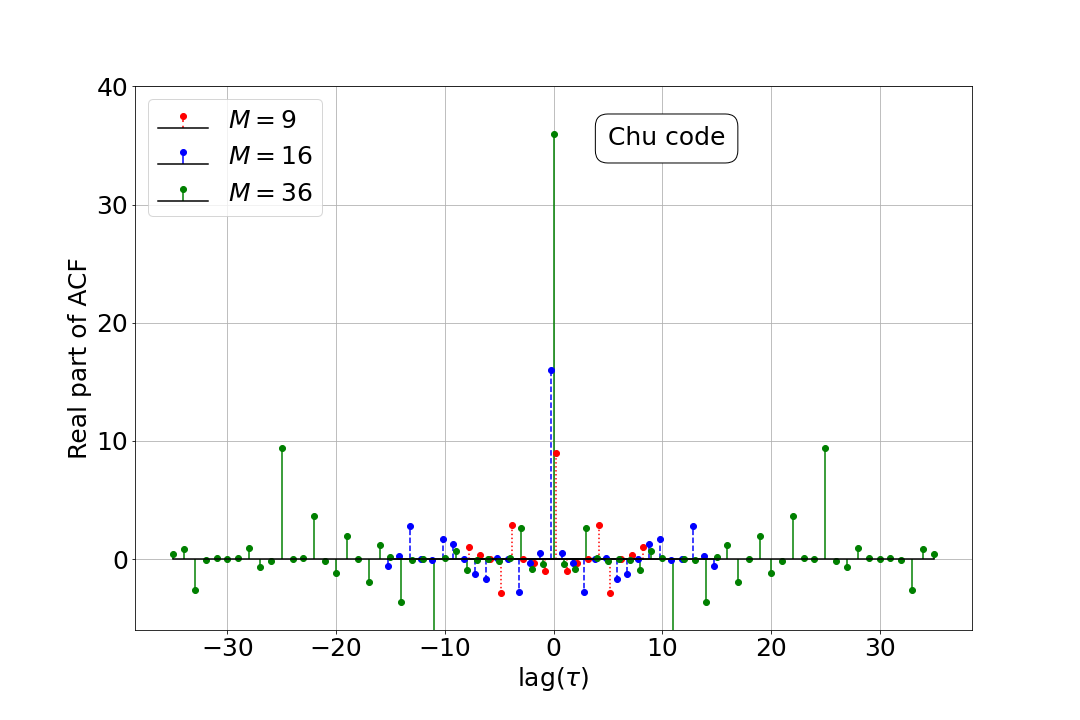}
        %\caption*{Chu code}
        %\label{fig:subfig3}
    \end{minipage}\hfill
    \begin{minipage}[b]{0.5\columnwidth}
        \centering
        \includegraphics[width=0.9\linewidth]{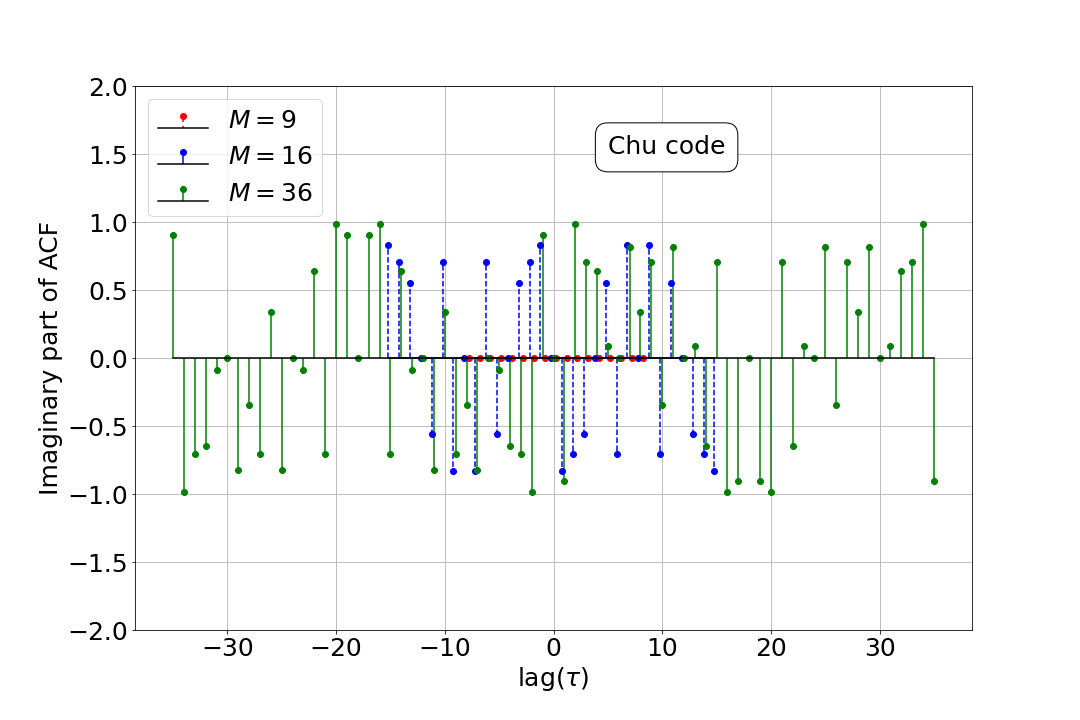}
        %\caption*{Chu code}
        %\label{fig:subfig4}
    \end{minipage}
    \caption{ACF of different codes.}
    \label{hovering energy}
\end{figure}

In Fig. \ref{hovering energy}, the real and imaginary parts of the ACF for the discussed codes are depicted for various values of \( M \) and different lag values. As it can be observed, for the Barker code, the imaginary part of the sidelobes in the ACF is zero, and the absolute value of real component is constrained to a maximum of one. When comparing the Frank code to the Chu code, it is evident that the Frank code exhibits a slightly higher imaginary part of the sidelobes in the ACF. However, the real part of the sidelobes in the Frank code is lower than in the Chu code. Overall, these codes are characterized by a delta-like ACF, where the sidelobe levels are markedly lower than the peak. Consequently, the resulting PDAF have a flat-like characteristic. In our simulations, the above-mentioned codes are employed to generate a flat-like PDAF.

\section{Spectral-efficient beamforming}\label{SE}
In this section, we derive the probabilistic lower and upper bounds on the average SE delivered to the UEs. These bounds are crucial for both the design and performance analysis of the proposed code.

We assume that $K$ UEs are uniformly distributed over a half-ring coverage area
\begin{equation}\label{halfregi}
\begin{cases}
          R_{1}\leq r \leq R_{2} \\\\
          \frac{-\pi}{2}\leq \theta \leq \frac{\pi}{2}. \\
     \end{cases}
\end{equation}
Defining $v := \frac{P}{\sigma^2} \beta_{h}(r_{h}) G_{0}(\theta_{h})$, the SE of the $k$-th UE is given by
\[S_{k}=\log_{2}\left(1+v\beta_{g}(r_{k}) G_{0}(\theta_{k})A(\mathbf{\Phi},\theta_{k})\right), \]
where $\beta_{h}(r_{h})$ and $\beta_{g}(r_{k})$ denote the path-loss terms at the reference RIS; $r_{h}$ and $r_{k}$ are the distances between the transmitter and the RIS and between the RIS and the $k$-th UE, respectively. Furthermore, $\theta_k$ is the AoD from the RIS to the $k$-th UE. We assume that the UEs are uniformly distributed such that $r_{k} \sim~U[R_{1},R_{2}]$, $\theta_k \sim U\left[-\frac{\pi}{2}, \frac{\pi}{2}\right]$, and $\theta_k$ is independent of $r_{k}$. To streamline our notation, let us define $G_{0}^{\text{min}} := \min\limits_{\theta \in \left[-\frac{\pi}{2}, \frac{\pi}{2}\right]} G_{0}(\theta)$, $G_{0}^{\text{max}} :=~\max\limits_{\theta \in \left[-\frac{\pi}{2}, \frac{\pi}{2}\right]} G_{0}(\theta)$, and $A^{\text{min}}(\mathbf{\Phi}) := \min\limits_{\theta \in \left[-\frac{\pi}{2}, \frac{\pi}{2}\right]} A(\mathbf{\Phi},\theta)$. In the following Proposition, we establish the probabilistic lower and upper bounds for $\overline{S}\triangleq \frac{1}{K}\sum\limits_{k=1}^{K}S_{k}$ which represents the average SE delivered to the UEs. 
\begin{proposition}\label{probound}
For any $\epsilon>0$, there exists a positive integer $N_{\epsilon}$ such that when $K\geq N_{\epsilon}$, the following holds with probability $1$ 
\begin{multline}\label{uplowb}
E_{r}\left\{\log_{2}\left(1+vG_{0}^{\text{min}} A^{\text{min}}(\mathbf{\Phi})\beta_{g}(r)\right)\right\}-\epsilon<\overline{S} \\
\overline{S}<\log_{2}\left(1+vG_{0}^{\text{max}}E_{r}\{\beta_{g}(r)\}E_{\theta}\{A(\mathbf{\Phi},\theta)\}\right)+\epsilon.
\end{multline}
\end{proposition}
\proof Please refer to Appendix \ref{app0a}.\qed

Although the bounds derived in the Proposition \ref{probound} may not be tight, they offer valuable insights into the key metrics that should be considered when designing the phase shifts of RIS elements. It can be noted that the lower bound increases with $A^{\text{min}}(\mathbf{\Phi})$. Thus, by addressing the optimization task $\max\limits_{\mathbf{\Phi}} A^{\text{min}}(\mathbf{\Phi})$, we can enhance this lower bound, leading to improved average SE. Similarly, the upper bound rises with $E_{\theta}\{A(\mathbf{\Phi},\theta)\}$. As such, a code with a higher value of $E_{\theta}\{A(\mathbf{\Phi},\theta)\}$ allows for a greater potential increase in average SE. 
\section{Broad Beamforming}\label{PS}
In this section, we introduce a novel code designed to optimize the phase shifts of RIS elements to achieve a flat-like PDAF. Motivated by the derived lower bound in Proposition \ref{probound}, this code maximizes the minimum value of the PDAF across all azimuth angles from \(\frac{-\pi}{2}\) to \(\frac{\pi}{2}\). Additionally, we derive a closed-form expansion for the average PDAF as a secondary metric for comparing the performance of different codes. As noted in Proposition \ref{probound}, a code that has higher value of $E_{\theta}\{A(\mathbf{\Phi},\theta)\}$ allows for a greater potential increase in average SE.

\subsection{Proposed Code}\label{PC}
Contrary to the previous codes that design phase shifts in the time domain by minimizing the sidelobes of the ACF, our proposed code seeks to design the phase shifts in the frequency domain, with the objective of rendering the PDAF flat-like. To accomplish this, we address the following optimization problem
\begin{equation}\label{asliopt}
\max_{\mathbf{\Phi}} \min_{\theta \in \left[-\frac{\pi}{2},\frac{\pi}{2}\right]} A(\mathbf{\Phi},\theta).
\end{equation}
Considering that the UEs are uniformly distributed around the RIS, we propose to maximize the worst case PDAF in problem (\ref{asliopt}). For a fixed phase shift vector $\mathbf{\Phi}$, the function $A(\mathbf{\Phi},\theta)$ is highly non-convex. Therefore, it is not possible to optimally solve the optimization problem $\min\limits_{\theta \in \left[-\frac{\pi}{2},\frac{\pi}{2}\right]} A(\mathbf{\Phi},\theta)$. To tackle this optimization problem, the continuous domain $\left[-\frac{\pi}{2},\frac{\pi}{2}\right]$ is discretized into a finite set of points defined as $\theta_i = \frac{-\pi}{2} + \frac{\pi i}{D}$, where $i=0, 1, 2, \ldots, D$. This discretization simplifies the problem by transforming an infinite set of possible values of $\theta$ into a finite set, making the problem more tractable computationally.

As $D$ increases, the resolution of the discretization improves, and the difference between adjacent discrete points, $\Delta \theta = \frac{\pi}{D}$, becomes smaller. This finer discretization allows for a more thorough exploration of the objective function $A(\mathbf{\Phi},\theta)$ within the given range. Hence, we use the following approximation 
\begin{equation}\label{appD}
\min_{\theta \in \left[-\frac{\pi}{2},\frac{\pi}{2}\right]} A(\mathbf{\Phi},\theta) \approx \min_{i=0,1,\ldots,D}A\left(\mathbf{\Phi},\frac{-\pi}{2}+\frac{\pi i}{D}\right),
\end{equation}
where $D$ is a large integer value. Let us define the approximation error $e$ as
\[e\triangleq \left|\min_{\theta \in \left[-\frac{\pi}{2},\frac{\pi}{2}\right]} A(\mathbf{\Phi},\theta) - \min_{i=0,1,\ldots,D}A\left(\mathbf{\Phi},\frac{-\pi}{2}+\frac{\pi i}{D}\right)\right |.\]

In what follows, we derive an asymptotic upper bound for the approximation error \(e\). To begin, we present a lemma that establishes an important property of the PDAF function.
\begin{lemma}\label{filem}
Given any phase shift vector $\mathbf{\Phi}$, the function $A(\mathbf{\Phi},\theta)$ is Lipschitz continuous\footnote{A function $f(\cdot)$ is called Lipschitz continuous on $[a,b]$ with Lipschitz constant $C$ if the inequality $|f(x_{2})-f(x_{1})|\leq C|x_{2}-x_{1}|$ holds for any $x_{1},x_{2} \in [a,b]$.} in $\theta$ with a Lipschitz constant of $\frac{(M-1)M^{2}\pi \Delta}{\lambda}$. Specifically, for any $\theta_{1}, \theta_{2} \in \left[-\frac{\pi}{2}, \frac{\pi}{2}\right]$, the following inequality holds
\begin{equation*}
\left|A(\mathbf{\Phi},\theta_{2}) - A(\mathbf{\Phi},\theta_{1})\right| \leq \frac{(M-1)M^{2}\pi \Delta}{\lambda} |\theta_{2} - \theta_{1}|.
\end{equation*}
\end{lemma}
\proof Please refer to Appendix \ref{appa}.\qed

We emphasize that Lipschitz continuity is a property that quantifies how smoothly a function behaves. Specifically, it provides a bound on how much the function can change when its input changes. The Lipschitz constant given in Lemma \ref{filem} is \( \frac{(M-1)M^{2}\pi \Delta}{\lambda} \). This constant depends on the number of RIS elements $M$ and ratio $\frac{\Delta}{\lambda}$. The form of this constant suggests that as the number of elements \( M \) increases or the ratio \( \frac{\Delta}{\lambda} \) becomes larger, the bound on how fast \( A(\mathbf{\Phi}, \theta) \) can change with \(\theta\) becomes larger. Lemma \(\ref{filem}\) provides the foundational basis necessary for proving the following theorem.
\begin{theorem}\label{thecom}
For a discretization with \( D+1 \) points, where \( D \) is sufficiently large\footnote{In the proof, we precisely specify how large \(D\) needs to be.}, we have
\begin{equation}\label{uppbon}
e<\frac{(M-1)M^{2}\pi^{2} \Delta}{\lambda D}.
\end{equation}
\end{theorem}
\proof Please refer to Appendix \ref{appb}.\qed

Although the obtained upper bound on the approximation error is not necessarily tight, it offers insights into the key parameters that influence the accuracy of the approximation. From the given upper bound (\ref{uppbon}), we observe that the approximation error decreases as \(D\) increases. Conversely, this upper bound increases with the number of elements \(M\), suggesting that higher values of \(M\) can lead to greater approximation errors. 
%To compensate for this error, it is necessary to proportionally increase the value of \(D\). 

By applying the discussed approximation, our goal optimization problem can be rewritten as
\begin{equation*}
\max_{\mathbf{\Phi}} \min_{i=0,1,\ldots,D }A\left(\mathbf{\Phi},\frac{-\pi}{2}+\frac{\pi i}{D}\right).
\end{equation*}

\begin{algorithm}[t]
 \begin{algorithmic}
 \renewcommand{\algorithmicrequire}{\textbf{Input: Fitness function $\eta(\boldsymbol{\theta},\boldsymbol{P})$ and population size $2L$}}
 %\REQUIRE 
%  \\ \textit{Initialisation} : $V^{(1)}=\{v_{i}\}$, $E^{(1)}=\{\}$ and $L=1$.
%   \STATE first statement
 %\\ %\textit{Phase 1}
 \STATE \textbullet\ Generate the initial population $\mathbb{P}^{(0)}= {\left[\mathbf{\Phi}_{1}^{(0)},\mathbf{\Phi}_{2}^{(0)},\ldots,\mathbf{\Phi}_{2L}^{(0)}\right]}^{T}$.
 \STATE \textbullet\ Let $\eta\left({\mathbf{\Phi}_{\ell}^{(0)}}\right)=\min\limits_{i=0,1,\ldots,D }A\left(\mathbf{\Phi}_{\ell}^{(0)},\frac{-\pi}{2}+\frac{\pi i}{D}\right)$ be the fitness value for $\ell=1,2,\ldots,2L$.
 \STATE \textbullet\ Define the selection random variable $Q^{(0)}$ on the set $\{1,2,\ldots,2L\}$ with probability mass function $Pr\{Q^{(0)}=k\}=\frac{\eta\left({\mathbf{\Phi}_{k}^{(0)}}\right)}{\sum\limits_{\ell=1}^{2L}\eta\left({\mathbf{\Phi}_{\ell}^{(0)}}\right)}$.
 \FOR {$i=1:N$}
 \STATE \textbullet\ Select \( 2L \) individuals from the set \( \mathbb{P}^{(i-1)} \) according to the distribution defined by the selection random variable \( Q^{(i-1)} \).
 \STATE \textbullet\ Randomly select \( L \) pairs of individuals from \( \mathbb{P}^{(i-1)} \) for crossover. Replace each selected pair with the two offsprings generated from their crossover.
 \STATE \textbullet\ Add a small random perturbation to each element of the generated offsprings (mutation).
 \STATE \textbullet\ Determine the updated population \( \mathbb{P}^{(i)} \) and the corresponding selection random variable \( Q^{(i)} \).
 \ENDFOR
 \end{algorithmic} 
 \caption{Proposed Code}
 \end{algorithm}
To address this optimization problem, we employ the CGA. The CGA is an optimization technique designed for finding optimal solutions within continuous parameter spaces \cite{Lit13}. It draws inspiration from the principles of natural selection and evolutionary processes, evolving a pool of potential solutions across successive generations to refine them progressively. The key phases of the CGA are as follows:
\begin{itemize}
\item \textbf{Initialization}: A population of initial candidate solutions is generated randomly, with the population size determined by the complexity of the problem. Each candidate is represented by a vector of continuous values.

\item \textbf{Selection}: Solutions are assessed using a fitness function that evaluates their performance on the problem. In this study, $\eta\left({\mathbf{\Phi} }\right)\triangleq \min\limits_{i=0,1,\ldots,D}A\left(\mathbf{\Phi},\frac{-\pi}{2}+\frac{\pi i}{D}\right)$ serves as the fitness function. The most effective solutions are chosen as parents for the subsequent generation.

\item \textbf{Crossover}: Offspring solutions are produced by combining pairs of parent solutions. This is achieved by calculating the weighted sum of the parent vectors, resulting in new solution vectors that embody characteristics from both parents. The crossover process aids in exploring the search space more comprehensively and generating a variety of candidate solutions.

\item \textbf{Mutation}: To diversify the search and avoid local optima, random alterations are made to the offspring solutions. This involves adding a small random value to each element of the solution vector. If the mutation step results in a variable falling outside the interval \([0, 2\pi]\), we automatically adjust the value to the nearest boundary. Specifically, if the mutation causes the value to drop below the lower bound \(0\), we reset it to \(0\). Conversely, if the value exceeds the upper bound \(2\pi\), we reset it to \(2\pi\). This clamping mechanism ensures that all variables remain within the defined range throughout the optimization process.
\end{itemize}
The CGA iterates through selection, crossover, and mutation until it meets a specified termination criterion. We provide detailed implementation of the CGA for optimizing the phase shifts of RIS elements in Algorithm 1. Finally, we point out that a rough estimate of the time complexity of a CGA can be expressed as $\mathcal{O}\left(N \times 2L \times (D+1) \times a\right)$, where $N$ is the number of generations, $2L$ is the population size, $D+1$ is the number of discrete points for approximating the value of $\min\limits_{\theta \in \left[-\frac{\pi}{2},\frac{\pi}{2}\right]} A(\mathbf{\Phi},\theta)$, and $a$ is the complexity of finding the value of $A(\mathbf{\Phi},\theta)$. While the time complexity of Algorithm 1 may initially appear high, it can be executed offline as it does not rely on real-time data. This allows for the optimized phase shifts to be precomputed and stored. When broad beamforming is needed, the RIS elements can be adjusted to these pre-stored values, ensuring efficient and rapid deployment.

\begin{figure}[t]
\centerline{\includegraphics[height=5.0cm,width=8.5cm]{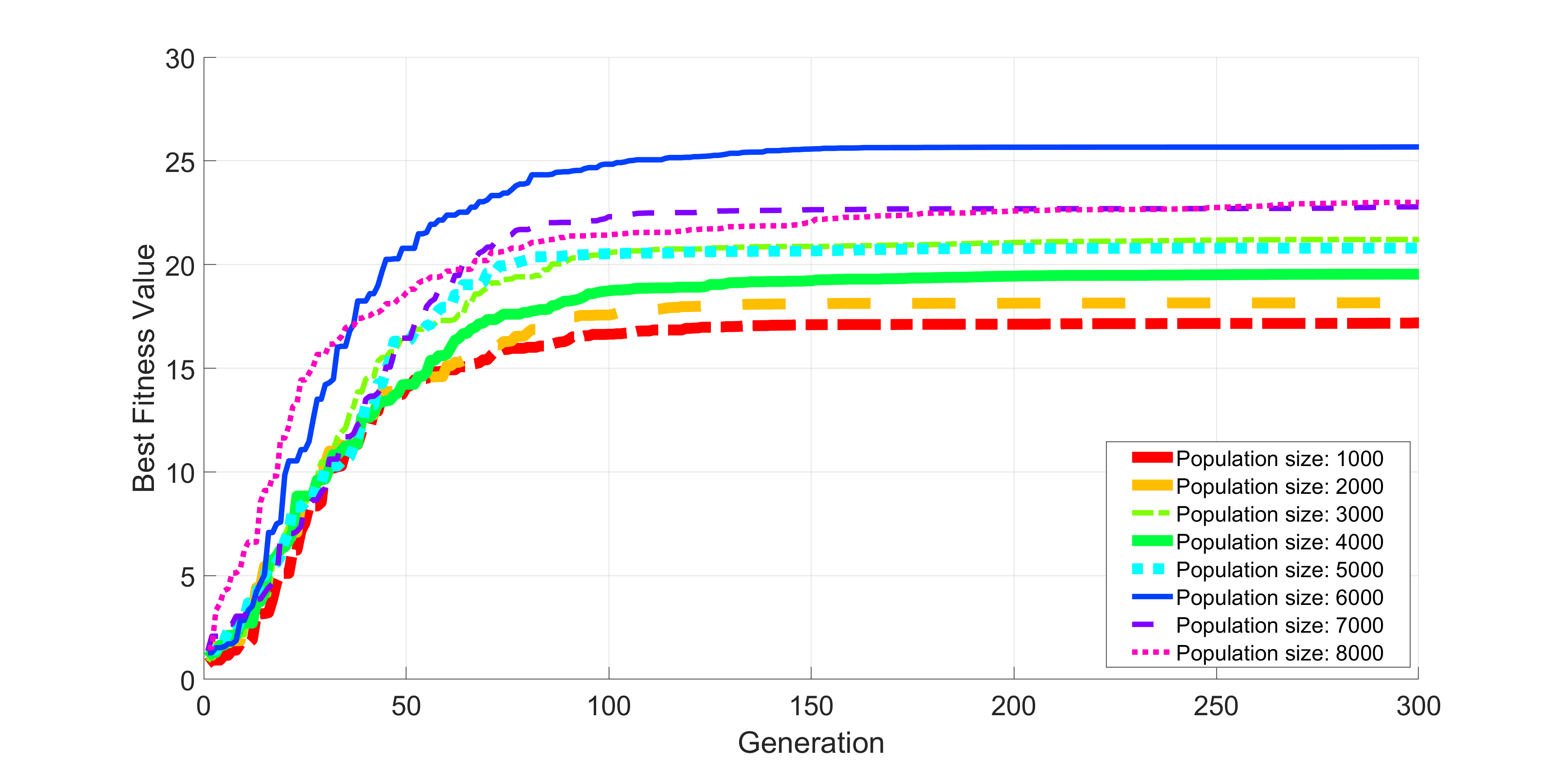}}
\caption{Convergence of Algorithm 1 for $M=64$.}
\label{convergence}
\end{figure} 
\begin{figure*}[b]
\hrule
\vspace{10pt} % Add some space between the line and the equation
\begin{equation}\label{expan}
U_{\frac{1}{2}} (\mathbf{\Phi})=\frac{M+2\sum\limits_{n=1}^{M-1}\sum\limits_{m=n+1}^{M}J_{0}((m-n)\pi)\cos{\left(\phi_{m}-\phi_{n}-(m-n)\pi\sin{\theta_{h}}\right)}}{M+2\sum\limits_{n=1}^{M-1}\sum\limits_{m=n+1}^{M}{(-1)}^{m-n}J_{0}((m-n)\pi)}
\end{equation}
\end{figure*}

\subsection{Second Metric: Average Value of PDAF}\label{SM}
So far, we have proposed a code for designing the phase shift of RIS elements with the aim of maximizing the minimum value of PDAF over azimuth angle $\theta \in \left[-\frac{\pi}{2},\frac{\pi}{2}\right]$. Comparing the average PDAF value across various codes is also important. Noting the derived upper bound on the average SE delivered to the UEs in Section \ref{SE}, which increases with $E_{\theta}\{A(\mathbf{\Phi},\theta)\}$, utilizing a code with a higher value of $E_{\theta}\{A(\mathbf{\Phi},\theta)\}$ allows for a greater potential increase in average SE. A closed-form expansion for $E_{\theta}\left\{A(\mathbf{\Phi},\theta)\right\}$ is given in the following theorem.
\begin{theorem}\label{therfin}
Assuming that $\theta \thicksim U\left[-\frac{\pi}{2},\frac{\pi}{2}\right]$, we have
\begin{equation*}
E_{\theta}\left\{A(\mathbf{\Phi},\theta)\right\}~~~~~~~~~~~~~~~~~~~~~~~~~~~~~~~~~~~~~~~~~~~~~~~~~
\end{equation*}
\begin{eqnarray*}
&=&2\sum_{n=1}^{M-1}\sum_{m=n+1}^{M}J_{0}(a_{m-n})\cos{\left(\phi_{m}-\phi_{n}-a_{m-n}\sin{\theta_{h}}\right)}\\
&~&+ M,
\end{eqnarray*}
where $J_{0}(\cdot)$ is the Bessel function of the first kind with order zero defined as
\begin{equation}\label{besseldef}
J_{0}(x)\triangleq E_{\theta} \left\{e^{jx  \sin{\theta} } \right\}=\frac{1}{\pi}\int_{-\frac{\pi}{2}}^{\frac{\pi}{2}} e^{jx  \sin{\theta} } \,d\theta,
\end{equation}
and $a_{k}$ is given by $a_{k}:=2\pi \frac{\Delta}{\lambda}k$ for integer values of $k$.
\end{theorem}
\proof Please refer to Appendix \ref{appcc}.\qed
%Please refer to Appendix \ref{appb}.\qed

As it can be observed, the value of $E_{\theta}\left\{A(\mathbf{\Phi},\theta)\right\}$ depends on the phase shift vector $\mathbf{\Phi}$ and ratio $\frac{\Delta}{\lambda}$. 
In this paper, we assume that $\frac{\Delta}{\lambda}=\frac{1}{2}$, which is more prevalent in the literature, especially in the field of antenna and array design \cite{antenref}. This spacing helps to avoid grating lobes (unwanted radiation peaks at certain angles) in the far-field radiation pattern of the array when it is steered away from the broadside direction. It also ensures that the array can achieve good performance in terms of beamforming and spatial resolution without unnecessarily increasing the complexity and cost. Hence, $\frac{\Delta}{\lambda}=\frac{1}{2}$ is considered a standard in many applications since it balances between performance and practicality. 

Now, let us define
\begin{equation}\label{defsef}
U_{\frac{\Delta}{\lambda}} (\mathbf{\Phi})\triangleq \frac{E_{\theta}\left\{A(\mathbf{\Phi},\theta)\right\}}{\max\limits_{\mathbf{\Phi}}E_{\theta}\left\{A(\mathbf{\Phi},\theta)\right\}}.
\end{equation}

Noting that \(0 < U_{\frac{\Delta}{\lambda}} (\mathbf{\Phi}) \leq 1\), \(U_{\frac{\Delta}{\lambda}} (\mathbf{\Phi})\) quantifies the normalized average PDAF value. Indeed, a high average PDAF value for a particular code implies that \(U_{\frac{\Delta}{\lambda}} (\mathbf{\Phi})\) is closer to $1$ for that code. We point out that this normalization facilitates easier comparisons between different codes.
In the following Proposition, we derive a closed-form expansion for the $U_{\frac{1}{2}} (\mathbf{\Phi})$.

\begin{proposition}\label{proppp2}
For a given phase shift vector $\mathbf{\Phi}$, the expanded formula for calculating $U_{\frac{1}{2}} (\mathbf{\Phi})$ is given in (\ref{expan}).
\end{proposition}
\proof Please refer to Appendix \ref{appddd}.\qed

In Section \ref{NR}, we consider $U_{\frac{1}{2}} (\mathbf{\Phi})$ as a metric for comparing the performance of different codes. 

\begin{table}[t] 
\centering % This centers the table on the page.
\caption{Table of the Proposed Code for Various $M$ values} % Caption for the table.
\label{listobtained}
\begin{tabular}{|c|c|} % Specifies that there are two centered columns; vertical bars indicate lines between and around the columns.
\hline % Inserts a horizontal line.
$M$ & Proposed Code ($\mathbf{\Phi}$)\\ \hline % The '&' character separates the columns, and '\\' starts a new row.
$13$ & \makecell{(0.9255,~4.6334,~2.0632,~5.6294,~3.0091,~3.3760,~0.8825,~1.4025,
\\
5.3366,~5.9881,~0.4434,~1.0334,~1.6318)} \\ \hline
$16$ & \makecell{(5.1194,~5.9698,~0.6334,~2.6752,~4.1800,~3.1756,~2.7309,~3.9677,
\\1.5603,~3.9362,~0.7873,~4.6435,~3.6637,~2.1817,~6.1171,~4.3546)} \\ \hline
$36$ & \makecell{(2.5458,~5.8091,~4.8406,~6.1005,~0.6850,~3.9223,~3.3962,~4.3150,
\\1.2969,~0.4361,~2.3980,~1.1877,~0.0314,~4.5914,~3.0137,~4.7279,
\\0.1981,~6.1708,~0.8647,~4.1516,~2.2116,~1.7623,~2.7874,~2.9865,
\\4.2146,~2.0956,~3.4289,~1.6303,~3.8954,~1.7463,~3.3191,~4.9514,
\\1.6416,~3.8338,~6.1454,~2.1315)}\\ \hline
$64$ & \makecell{(6.2154,~0.9377,~5.9622,~2.3982,~2.0550,~5.2374,~4.6543,~6.1568,
\\2.0360,~3.3983,~3.2310,~4.6648,~5.7030,~0.2454,~1.3573,~5.7993,
\\4.0525,~1.7875,~2.6875,~4.6555,~4.1909,~3.3503,~4.3216,~3.1219,
\\4.1523,~1.4800,~3.9280,~1.8255,~2.5134,~4.4152,~1.4448,~2.7278, 
\\6.0119,~5.6658,~3.4961,~3.4214,~1.3725,~3.1275,~0.9327,~3.4367,
\\4.3562,~5.5590,~3.9093,~3.5500,~4.7366,~3.0102,~0.2376,~4.6770,
\\5.2783,~2.3078,~0.3114,~4.7210,~0.4638,~5.7200,~4.5785,~1.8969,
\\5.5300,~1.3563,~5.1408,~4.0969,~1.8617,~0.9126,~5.7912,~4.1614)} \\ \hline
% Add more rows as needed.
\end{tabular}

\label{tab:example} % Label for referencing the table in text.
\end{table}

\section{Numerical Results}\label{NR}

In this section, we present numerical results to assess the
effectiveness of the proposed code for 
broad beamforming. In the simulation, we assume that $\lambda=10$ cm, $\Delta=\frac{\lambda}{2}$, $\theta_{h}=0$, and $M \in~\{13,16,36,64\}$.

\begin{figure*}[t]  % Use figure* for spanning across two columns in a two-column document
    \centering
    \begin{minipage}{0.5\textwidth}
        \centering
        \includegraphics[width=0.9\linewidth]{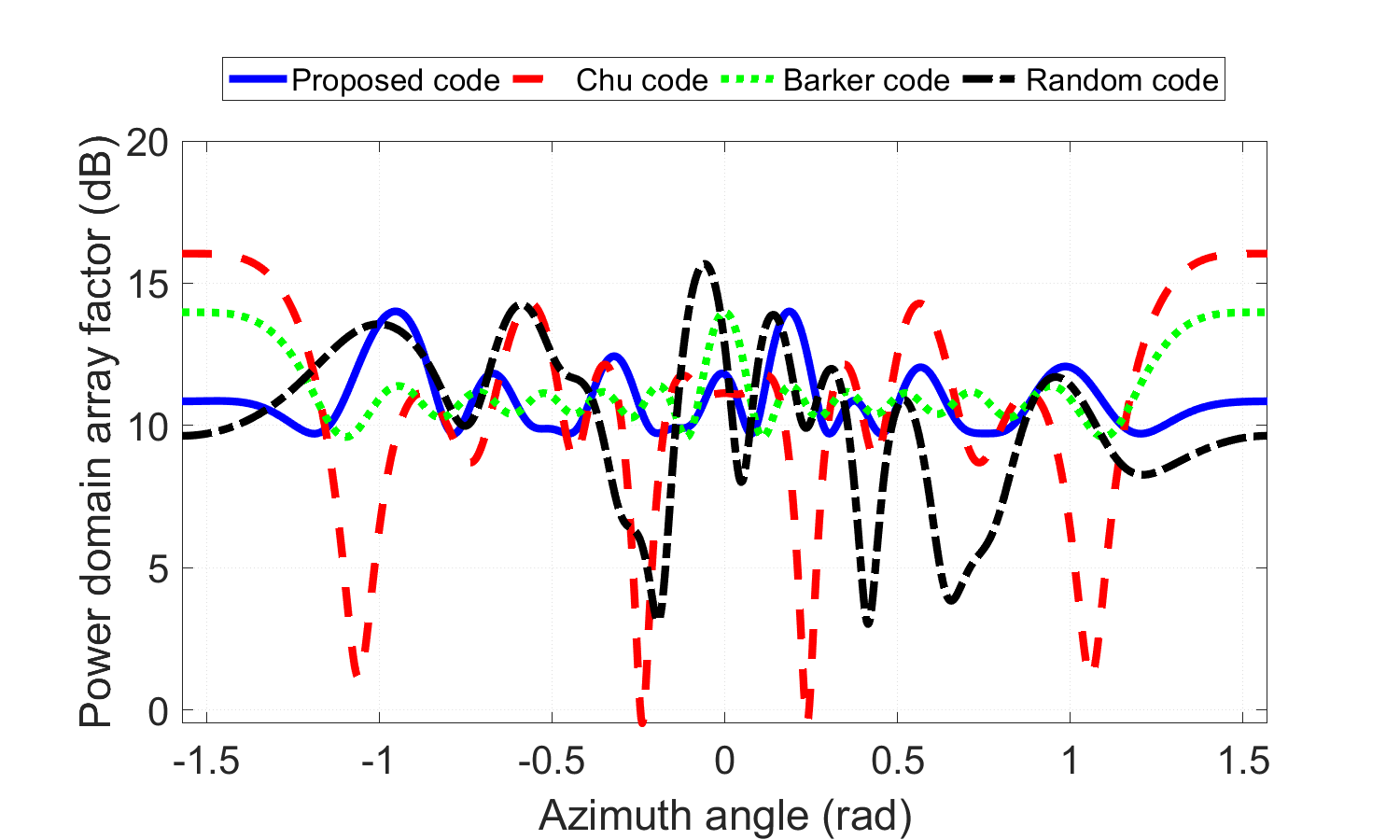}
        \caption*{(a) $M=13$}  % Optional: add subcaption for clarity
        %\label{fig:subfig1}
    \end{minipage}\hfill
    \begin{minipage}{0.5\textwidth}
        \centering
        \includegraphics[width=0.9\linewidth]{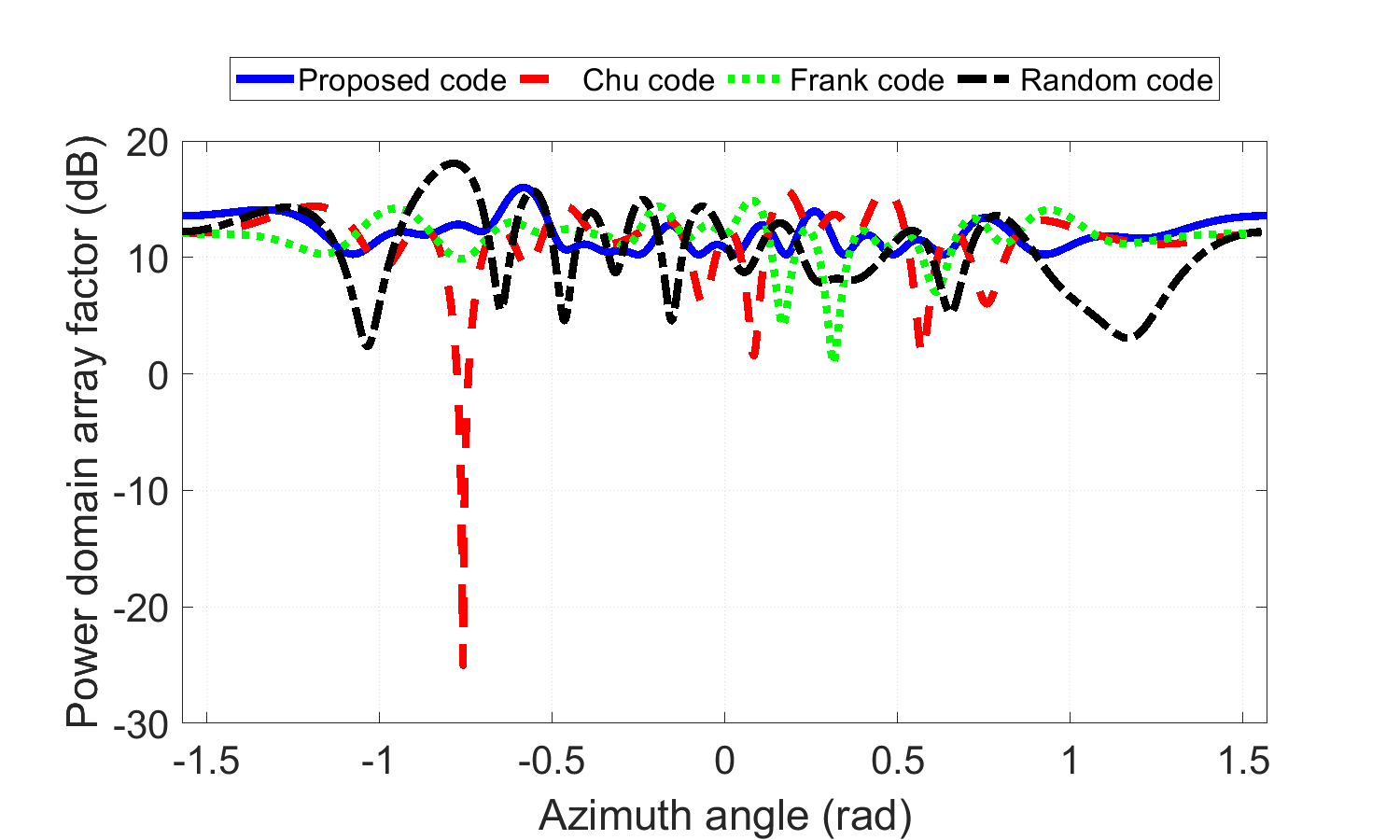}
        \caption*{(b) $M=16$}  % Optional: add subcaption for clarity
        %\label{fig:subfig2}
    \end{minipage}

    \vspace{0.5cm} % Optional: Adds vertical space between the rows of figures

    \begin{minipage}{0.5\textwidth}
        \centering
        \includegraphics[width=0.9\linewidth]{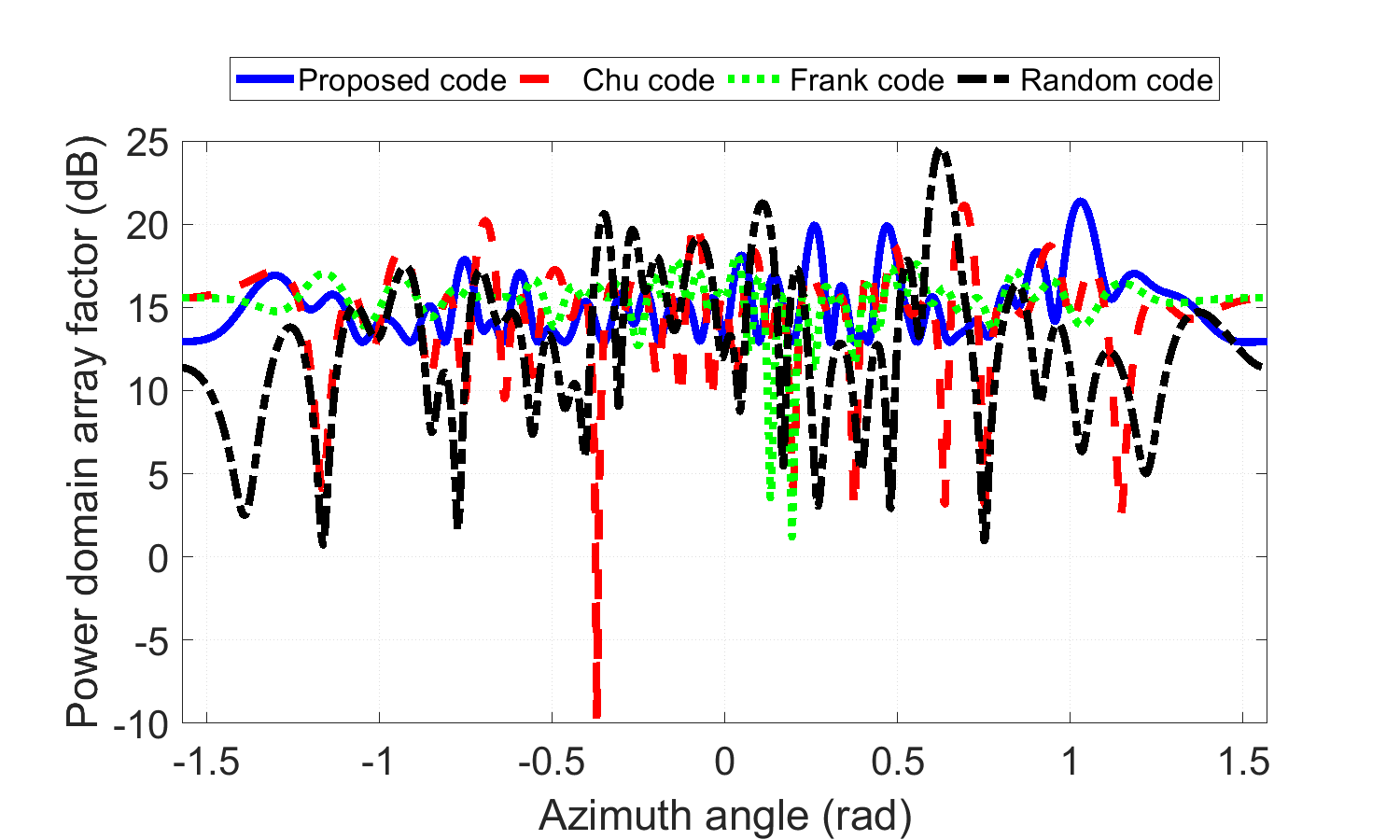}
        \caption*{(c) $M=36$}  % Optional: add subcaption for clarity
        %\label{fig:subfig3}
    \end{minipage}\hfill
    \begin{minipage}{0.5\textwidth}
        \centering
        \includegraphics[width=0.9\linewidth]{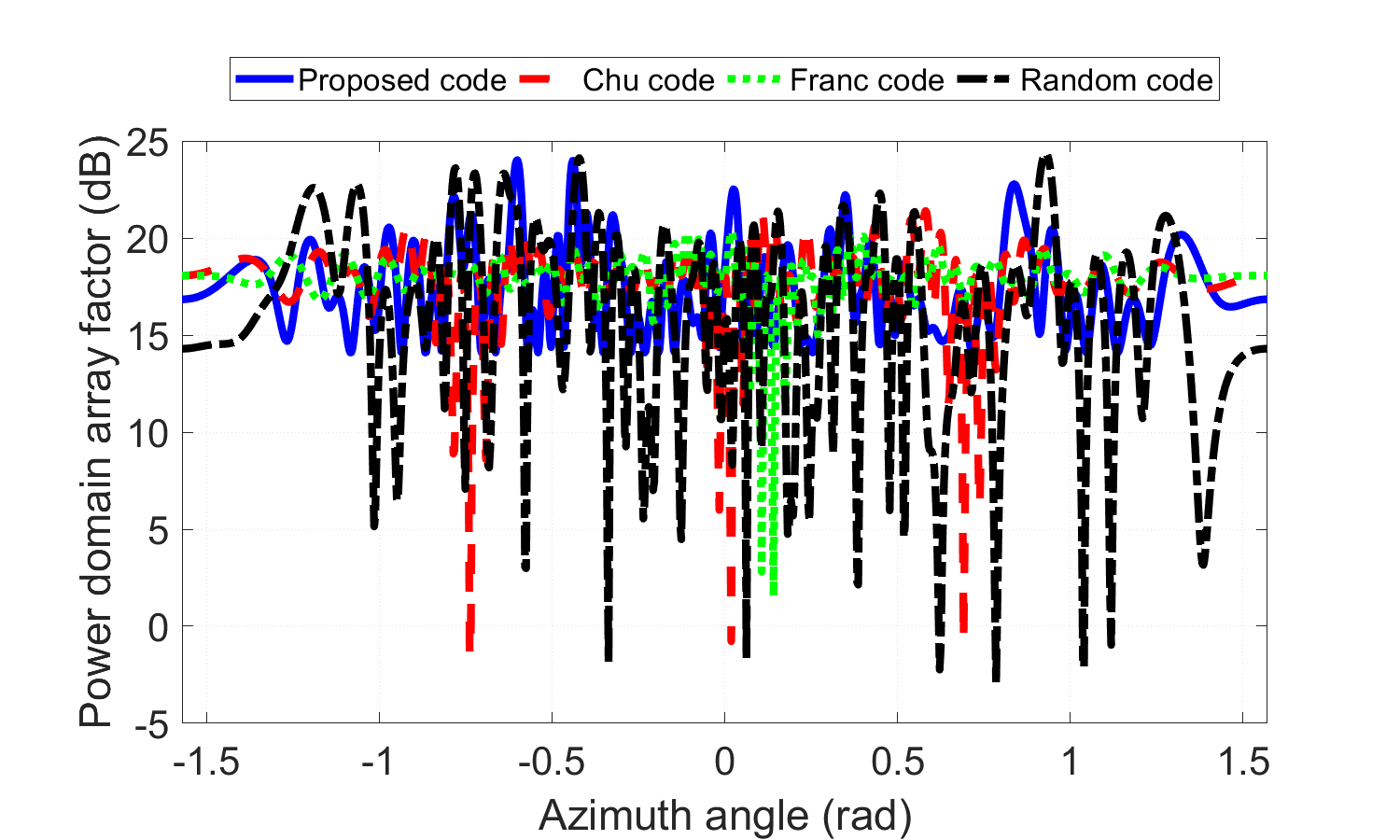}
        \caption*{(d) $M=64$}  % Optional: add subcaption for clarity
        %\label{fig:subfig4}
    \end{minipage}
    \caption{PDAF versus azimuth angle for various $M$ values.}
    \label{fig:hovering energy}
\end{figure*}
\begin{table}[t]
\centering
\caption{Obtained Numerical Values for the PDAF}
\label{simres}
\begin{tabular}{ccccc}
\toprule
\multicolumn{1}{c}{} & \multicolumn{2}{c}{\textbf{$M=13$}} &  \multicolumn{2}{c}{\textbf{$M=16$}}\\
\cmidrule(rl){2-3}  \cmidrule(rl){4-5}
\textbf{Codes} & {$A^{\text{min}}(\mathbf{\Phi})$} dB& {$U_{\frac{1}{2}} (\mathbf{\Phi})$} & {$A^{\text{min}}(\mathbf{\Phi})$} dB& {$U_{\frac{1}{2}} (\mathbf{\Phi})$} \\
\midrule
Proposed code  & $9.7142$  &  $0.3181$  & $10.2373$  &  $0.3103$  \\
Chu code &  $-0.4627$ & $0.4168$  &  $-25.0169$ & $0.2941$\\
Barker code & $9.5994$  &  $0.3634$  &  $-$ & $-$ \\
Frank code & $-$ & $-$ & $0.9454$  &  $0.2907$ \\
%Barker code  &  $-$ & $-$  & $-$  \\
random code  &  $3.0211$ & $0.3086$  &  $2.3690$ & $0.2891$ \\
%Golay code & $-$ & $-$  & $12.0412$ & $0.2919$ \\
%maximum-average code& $-66.6265$ & $1$   &   &   &   &  \\
%Golay code  &  $12.0412$ & $0.2919$ & $0.5520$ & $2.8428$ &  $\left(2.8428-0.0021,2.8428+0.0021\right)$\\
\bottomrule
\end{tabular}
\end{table} 
\begin{table}[t]
\centering
\begin{tabular}{ccccc}
\toprule
\multicolumn{1}{c}{} & \multicolumn{2}{c}{\textbf{$M=36$}} &  \multicolumn{2}{c}{\textbf{$M=64$}}\\
\cmidrule(rl){2-3}  \cmidrule(rl){4-5}
\textbf{Codes} & {$A^{\text{min}}(\mathbf{\Phi})$} dB& {$U_{\frac{1}{2}} (\mathbf{\Phi})$} & {$A^{\text{min}}(\mathbf{\Phi})$} dB& {$U_{\frac{1}{2}} (\mathbf{\Phi})$} \\
\midrule
Proposed code  & $12.9047$  &  $0.1933$   & $14.0971$  &  $0.1437$  \\
Chu code&  $-9.8148$ & $0.1939$ &  $-1.6166$ & $0.1471$ \\
Frank code & $1.2058$  &  $0.1959$ & $1.5626$  &  $0.1472$  \\
%Barker code  &  $-$ & $-$  & $-$  \\
random code  &  $0.7339$ & $0.1643$  &  $-2.9884$ & $0.1414$\\
%maximum-average code& $-66.6265$ & $1$   &   &   &   &  \\
%Golay code  &  $15.5630$ & $0.1956$  &  $18.0618$ & $0.1470$\\
\bottomrule
\end{tabular}
\end{table} 
For designing the phase shifts of RIS elements, we consider the following codes.
\begin{itemize}
    \item \textbf{Proposed code}: We employ Algorithm 1 that utilizes CGA  to design the code. Initially, we set \( D \) to $1000$ in the approximation formula given in (\ref{appD}). We then test various population sizes, setting \( 2L = 1000\ell \) for \( \ell = 1, 2, \ldots, 8 \). The algorithm is executed $50$ times for each population size. We also set the maximum number of generations to $300$. Ultimately, we select the best solution based on the outcomes from the different population sizes and runs. In Fig. \ref{convergence}, the best fitness values obtained by Algorithm 1 are plotted against the generation index for various population sizes when \( M=64 \). As it can be observed, the best fitness value is achieved when the population size ($2L$) is set to $6000$. The list of proposed codes obtained for various values of \(M\) is presented in Table \ref{listobtained}. Finally, we note that although these codes are derived when \(\theta_h=0\), they can be readily updated for a general \(\theta_h\) without the need to re-execute the CGA. To illustrate, if \(\phi_m\) represents the phase shift of the \(m\)-th RIS element for the case of \(\theta_h=0\), the adjusted phase shift for a general \(\theta_h\) is 
    \[\phi_m + \frac{2\pi \Delta (m-1)}{\lambda} \sin(\theta_h)~(\text{mod}~ 2\pi).\]
    \item \textbf{Chu code}: This code is defined in (\ref{autfunn}) to design the phase shifts of the RIS elements. We optimize the parameter \(q\) in (\ref{autfunn}) to maximize \(A^{\text{min}}(\mathbf{\Phi})\), and the optimal values obtained are \(3\), \(11\), \(13\), and \(43\) for \(M=~13, 16, 36, \text{and}~ 64\), respectively. We point out that these optimized values are relatively prime to their corresponding \(M\) values.
    \item \textbf{Frank code}: We employ the Frank code as specified in (\ref{Francformul}). Note that this code is only defined when \( M \) is a perfect square integer as mentioned in Section \ref{Flat}.
    \item \textbf{Barker code}: We use the Barker code with length $13$ as given in Table \ref{Barkcod}.
    \item \textbf{Random code}: We generate $1000$ random phase shift vectors and select the one that yields the maximum value of \( A^{\text{min}}(\mathbf{\Phi}) \).  
\end{itemize}

Fig. \ref{fig:hovering energy} displays the PDAF of various coding methods for \(M \in \{13,16,36,64\}\). As anticipated by Lemma \ref{filem}, the PDAF exhibits a more rapid variation with respect to $\theta$ as the value of $M$ increases. In addition, the  minimum and the normalized average PDAF values of the different coding methods are presented in Table \ref{simres} for easy comparison.

\begin{figure*}[t]  % Use figure* to span across two columns
    \centering
    \begin{minipage}[b]{0.5\textwidth}
        \centering
        \includegraphics[width=0.9\linewidth]{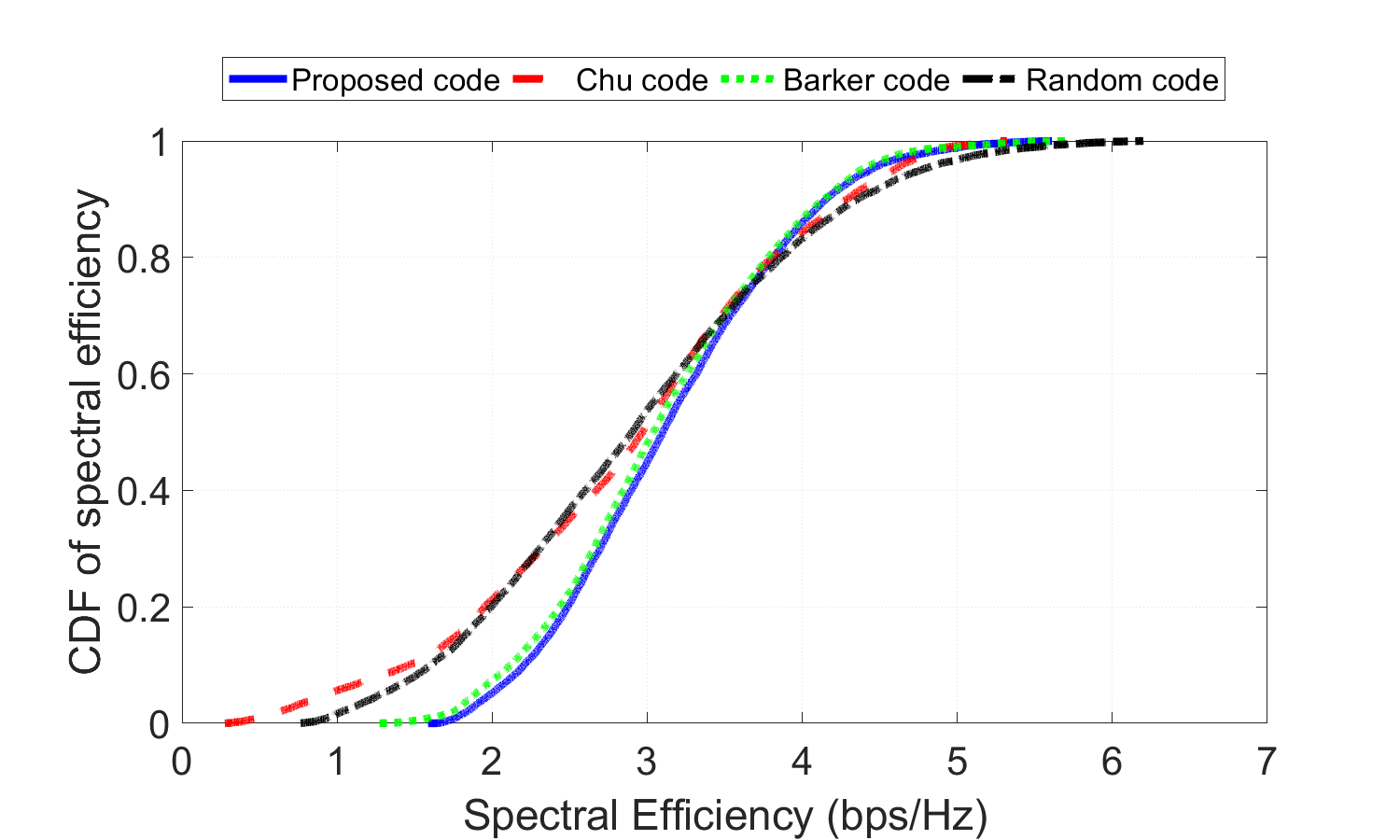}
        \caption*{(a) $M=13$}  % Optional: add subcaption for clarity
        %\label{fig:subfig1}
    \end{minipage}\hfill
    \begin{minipage}[b]{0.5\textwidth}
        \centering
        \includegraphics[width=0.9\linewidth]{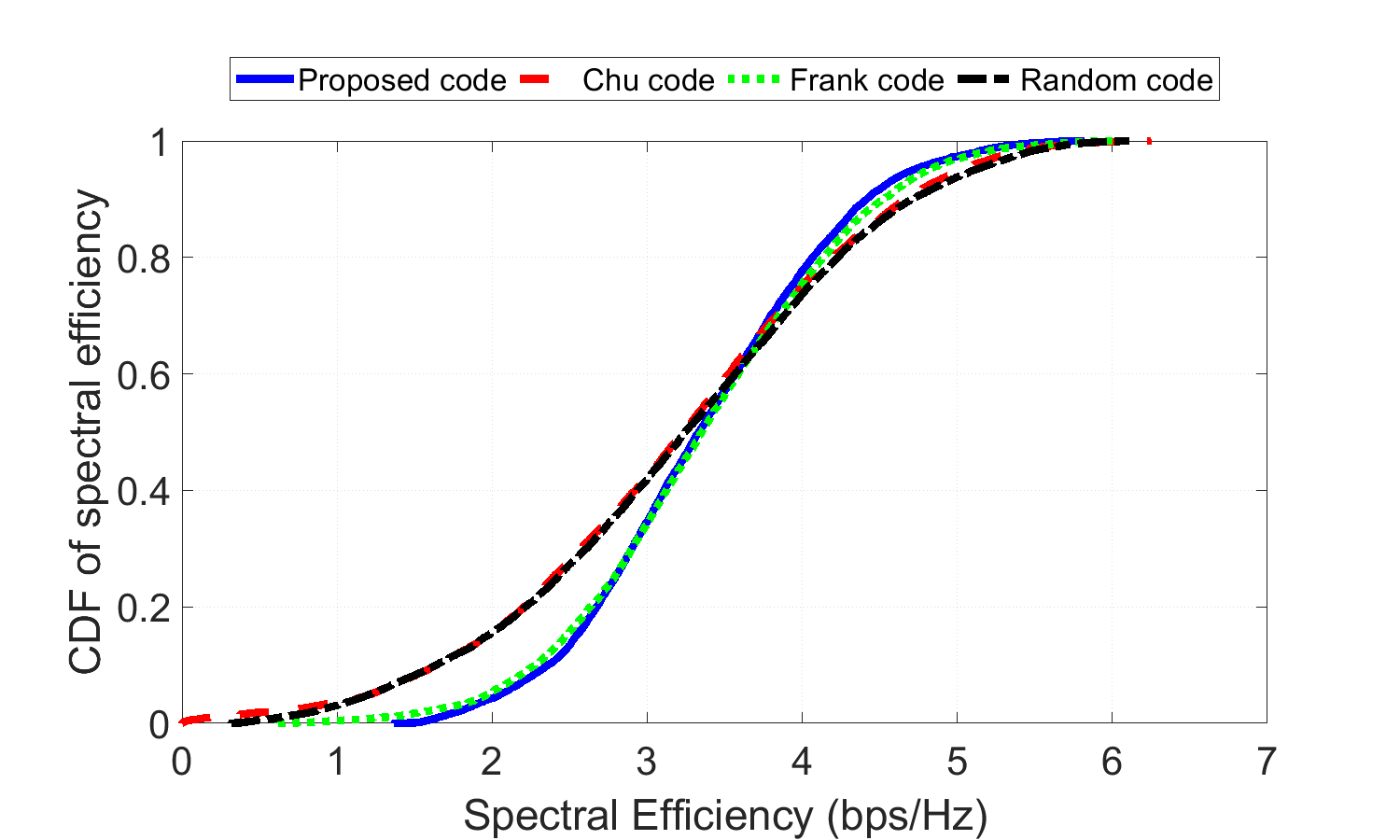}
        \caption*{(b) $M=16$}  % Optional: add subcaption for clarity
        %\label{fig:subfig2}
    \end{minipage}

    \vspace{0.5cm} % Optional: Adds vertical space between the rows of figures

    \begin{minipage}[b]{0.5\textwidth}
        \centering
        \includegraphics[width=0.9\linewidth]{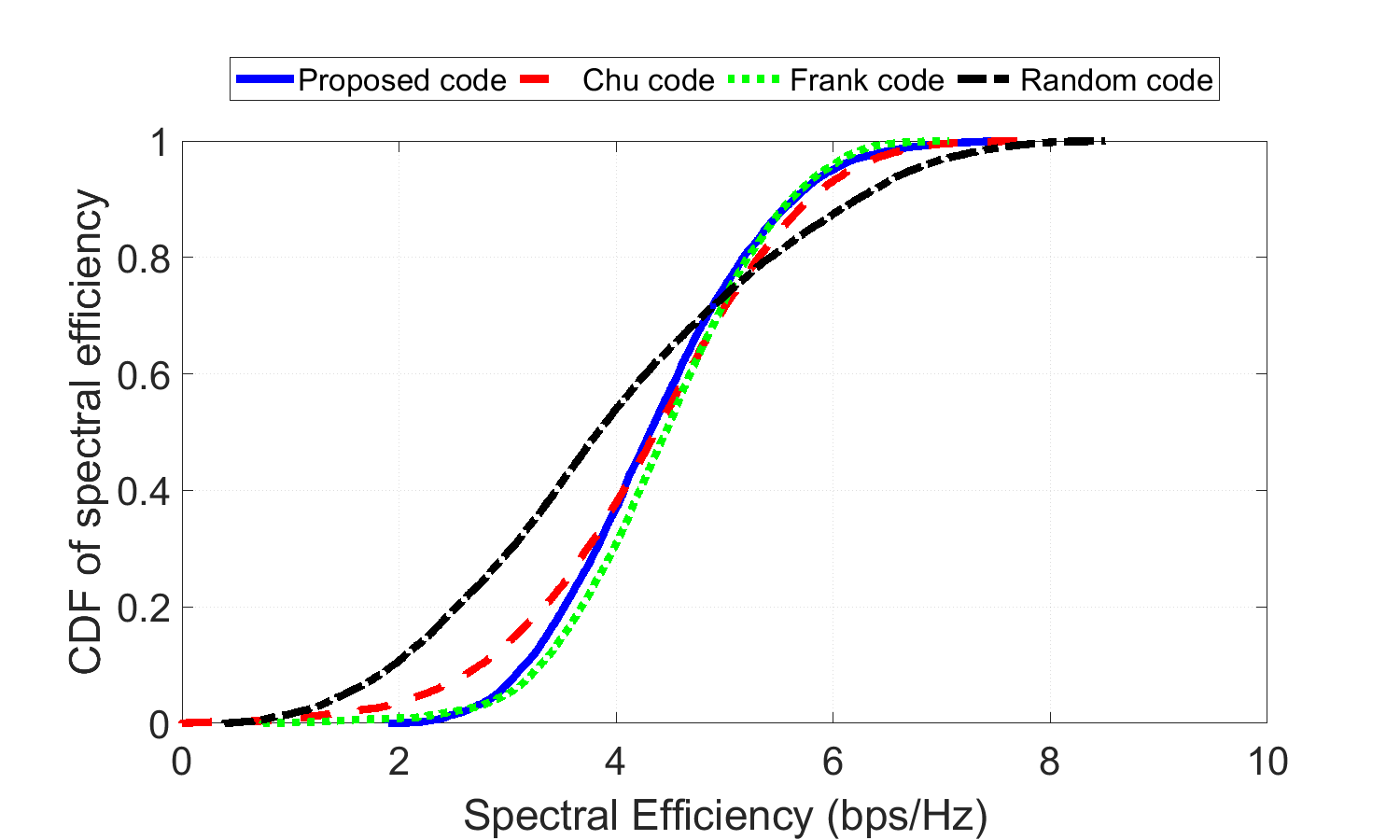}
        \caption*{(c) $M=36$ }  % Optional: add subcaption for clarity
        %\label{fig:subfig3}
    \end{minipage}\hfill
    \begin{minipage}[b]{0.5\textwidth}
        \centering
        \includegraphics[width=0.9\linewidth]{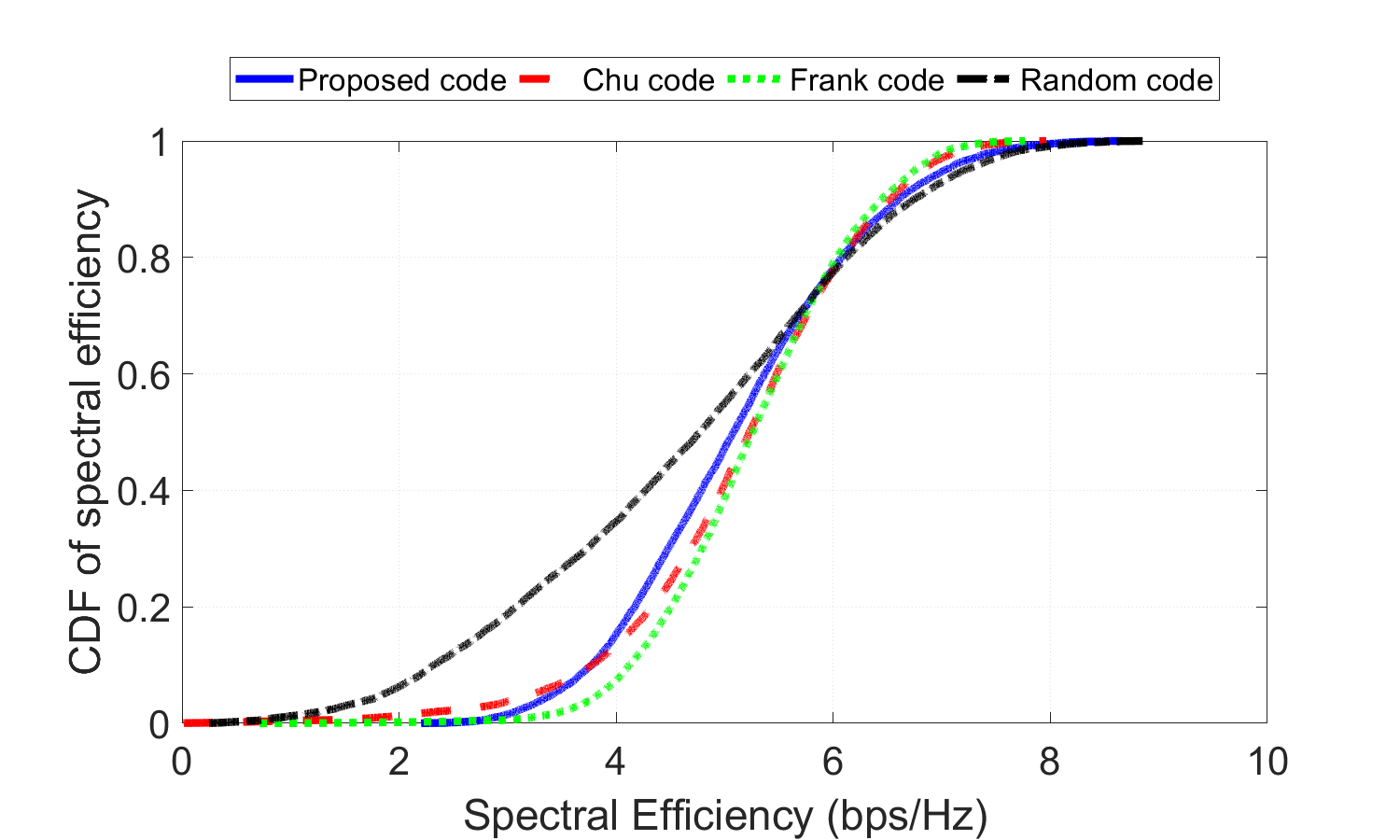}
        \caption*{(d)  $M=64$}  % Optional: add subcaption for clarity
        %\label{fig:subfig4}
    \end{minipage}
    \caption{Empirical CDF of spectral efficiency.}
    \label{ecdf}
\end{figure*} 

For \(M=13\), the Frank code is not applicable because $13$ is not a perfect square. Consequently, we compare the proposed code with the Barker, Chu, and random codes only when $M=13$. As can be observed from Fig. \ref{fig:hovering energy}, both the proposed code and the Barker code are effective in designing a beam with a flat-like PDAF. 
%In Table \ref{simres}, the values of the metrics considered are presented for various coding methods and \(M\) values.
As can be seen from Table \ref{simres}, the minimum array factor, \(A^{\text{min}}(\mathbf{\Phi})\), is higher for the proposed code compared to the Barker code. This difference arises because the phase shift values in the Barker code are restricted to either \(0\) or \(\pi\), whereas the phase shift values of the proposed code can vary across the full range from \(0\) to \(2\pi\). 
%We also compare the average value of the PDAF for different coding methods in Table \ref{simres}. 
It is evident that the Chu code exhibits the highest value of \(U_{\frac{1}{2}} (\mathbf{\Phi})\). Following this, the Barker code has the second highest value, succeeded by the proposed code, and lastly, the random code.
%%%%%%%%%%%%%%%%%%%%%%%%%%%%%%%%%%%%%%%%%%%%%%%%%%%%%%%%%%%%%%%%%%%%%%%%%%%%%

%%%%%%%%%%%%%%%%%%%%%%%%%%%%%%%%%%%%%%%%%%%%%%%%%%%%%%%%%%%%%%%%%%%%%%%%%%%

For \(M \in \{16,36,64\}\), we evaluate the performance of the proposed code against the Chu, Frank, and random codes. As it can be seen in Fig. \ref{fig:hovering energy} and Table \ref{simres}, \(A^{\text{min}}(\mathbf{\Phi})\) is significantly higher compared to those of the Chu, Frank, and random codes. For an instance, for the case of $M=64$, the values of $A^{\text{min}}(\mathbf{\Phi})$ are $14.0971$ dB, $1.5626$ dB, $-1.6166$ dB, and $-2.9884$ dB for the proposed code, Frank code, Chu code, and random code, respectively. As shown in Table \ref{simres}, \(U_{\frac{1}{2}} (\mathbf{\Phi})\) is higher for the proposed code compared to the other codes when $M=16$. For $M=36$ and $M=64$, \(U_{\frac{1}{2}} (\mathbf{\Phi})\) remains fairly consistent across the proposed code, Frank code, and Chu code. In summary, the discussion above concludes that the proposed code generates a superior broad beam compared to other well-known codes designed for uni-polarized RISs. Therefore, the proposed code is highly suitable for applications requiring broad beam where a uni-polarized RIS is employed.

\begin{table*}[t]
\centering
\caption{Obtained Numerical Values for the SE}
\label{seres}
\begin{tabular}{ccccc}
\toprule
\multicolumn{1}{c}{} & \multicolumn{2}{c}{\textbf{$M=13$}} & \multicolumn{2}{c}{\textbf{$M=16$}} \\
\cmidrule(rl){2-3}  \cmidrule(rl){4-5}
\textbf{Codes}  & $S^{\text{min}}~(\text{bps}/\text{Hz}) $ & $\overline{S}~(\text{bps}/\text{Hz}) $  & $S^{\text{min}}~(\text{bps}/\text{Hz}) $ & $\overline{S}~(\text{bps}/\text{Hz}) $ \\
\midrule
Proposed code  &  $1.5870$ & $3.1530\substack{+ \\ -} 0.0146$   &  $1.3668$ & $3.3671\substack{+ \\ -}0.0157$\\
Chu code &  $0.2764$ & $2.8877\substack{+ \\ -}0.0207$  &  $7.1792\times 10^{-5}$ & $3.1795\substack{+ \\ -}0.0228$\\
Barker code &  $1.2734$ & $3.0983\substack{+ \\ -}0.0150$  &  $-$ & $-$ \\

Frank code &  $-$ & $-$ & $0.6173$ & $3.3747\substack{+ \\ -}0.0171$ \\
random code  &  $0.7650$ & $2.9459\substack{+ \\ -}0.0206$  &  $0.3154$ & $3.2189\substack{+ \\ -}0.0226$\\

\bottomrule
\end{tabular}
\end{table*} 
\begin{table*}[t]
\centering
\begin{tabular}{cccccc}
\toprule
\multicolumn{1}{c}{} & \multicolumn{2}{c}{\textbf{$M=36$}} & \multicolumn{2}{c}{\textbf{$M=64$}} \\
\cmidrule(rl){2-3}  \cmidrule(rl){4-5}
\textbf{Codes}  & $S^{\text{min}}~(\text{bps}/\text{Hz}) $ & $\overline{S}~(\text{bps}/\text{Hz}) $  & $S^{\text{min}}~(\text{bps}/\text{Hz}) $ & $\overline{S}~(\text{bps}/\text{Hz}) $ \\
\midrule
Proposed code  &  $1.8994$ & $4.3633\substack{+ \\ -}0.0186$  &  $2.2025$ & $5.1303\substack{+ \\ -}0.0214$ \\
Chu code &  $1.1775\times 10^{-5}$ & $4.3008\substack{+ \\ -}0.0235$ &  $0.0140$ & $5.1536\substack{+ \\ -}0.0219$  \\
Frank code &  $0.7432$ & $4.4398\substack{+ \\ -}0.0178$  & $0.7140$ & $5.2631\substack{+ \\ -}0.0175$   \\

random code  & $0.3896$ & $3.9691\substack{+ \\ -}0.0312$  & $0.2507$ & $4.6423\substack{+ \\ -}0.0325$ \\

\bottomrule
\end{tabular}
\end{table*} 
\subsection{MCMC Simulation for Estimating $E_{r,\theta}\{S\}$}
To compare the achieved spectral-efficiency of different coding methods, we consider a RIS-assisted downlink setting with $K=10000$ UEs. The SE of the $k$-th UE is given by $S_k = \log_2\left(1 +  v\beta_g(r_k)  G_0(\theta_k) A(\mathbf{\Phi}, \theta_k)\right)$, where $v=~\frac{P}{\sigma^2} \beta_{h}(r_{h}) G_{0}(\theta_{h})$, and the radiation pattern of each element is obtained using the 3GPP model \cite{ramezani}, which is given by
\begin{equation*}
G_{0}(\theta)=8-\min{\left({12\left(\frac{\theta-\theta_{0}}{\Delta\theta}\right)}^{2},30\right)} (\text{dBi}),
\end{equation*}
with $\theta_{0}=0$ and $\Delta \theta=\frac{\pi}{2}$. Moreover, the path-loss is modeled as 
\[\beta_{h}(r_{h})=-37.5-22\log_{10}\left(r_{h}\right)\text{(dB)}\] 
and 
\[\beta_{g}(r_{k})=-37.5-22\log_{10}\left(r_{k}\right)\text{(dB)}.\]  
We also assume that $r_h = 50$~m and the UEs are uniformly distributed around the RIS such that $r_{k} \sim~U[50 \, \text{m}, 100 \, \text{m}]$ and $\theta_k \sim U\left[-\frac{\pi}{3}, \frac{\pi}{3}\right]$. The transmit power is set to $P = 47$ dBm, and the noise power is given by $\sigma^2 = -90$ dBm.

Since deriving a closed-form expression for the true expected value $E_{r,\theta}\{S\}$ is infeasible, we take advantage of MCMC simulations to approximate $E_{r,\theta}\{S\}$. Each iteration of the MCMC simulation generates a set of values for \(r_k\) and \(\theta_k\) based on their respective distributions. Using these generated values, \(S_k\) is computed for all \(k\). The MCMC procedure is repeated for a large number of iterations to ensure a thorough exploration of the parameter space. The sample average of $SE$ is given by $\overline{S}\triangleq \frac{1}{K}\sum\limits_{k=1}^{K}S_{k}$, which provides an approximation for the true expected value $E_{r,\theta}\{S\}$. To assess the accuracy of our approximation, we calculate an approximate 95\% confidence interval for $E_{r,\theta}\{S\}$ using the standard error (se) of MCMC samples, which is defined as \cite{montebook}
\[
se=\frac{1}{\sqrt{K}}\sqrt{\frac{1}{K-1}\sum\limits_{k=1}^{K}{\left(S_{k}-\overline{S}\right)}^{2}}.
\]
The 95\% confidence interval is then approximated as \cite{montebook}
\[
\text{CI} = \left(\overline{S} - 1.96 \times se, \overline{S} + 1.96 \times se\right).
\]
This interval provides an estimate of the range within which the true expected value of SE, i.e., $E_{r,\theta}\{S\}$ is likely to lie, assuming the normality of the distribution of $\overline{S}$, which is justified by the central limit theorem (CLT) given a large enough sample size $K$.

In Fig. \ref{ecdf}, the empirical CDF of the SE delivered by different codes to the UEs is presented. For an accurate comparison between codes, the values of the minimum SE, denoted as $S^{\text{min}}$, and the average SE (\(\overline{S}\)) are also presented in Table \ref{seres}. Specifically, for \(M=13\), the average SE values are \(3.1530\) bps/Hz for the proposed code, \(3.0983\) bps/Hz for the Barker code, \(2.9459\) bps/Hz for the random code, and \(2.8877\) bps/Hz for the Chu code. A similar pattern is observed for the minimum value of SE delivered to the UEs, indicating that the proposed code outperforms the Barker code in terms of SE. As discussed earlier, the value of \(A^{\text{min}}(\mathbf{\Phi})\) is higher for the proposed code compared to the Barker code, suggesting that the proposed code is more effective in generating a broad and spectral efficient beam. For \(M=16\) and \(M=36\), while the average SE value of the Frank code is marginally higher than that of the proposed code, the proposed code significantly outperforms the Frank code in terms of minimum SE. For \(M=64\), the minimum SE ($S^{\text{min}}$) of the proposed code is approximately three times greater than that of the Frank code, but the average SE (\(\overline{S}\)) is slightly higher for both the Frank and Chu codes compared to the proposed code.

In summary, the conducted simulation results clearly demonstrate that the proposed code outperforms the other codes in terms of the minimum PDAF (\(A^{\text{min}}(\mathbf{\Phi})\)) and the minimum SE (\(S^{\text{min}}\)) delivered to the UEs for various values of \(M\). This significant improvement in the minimum values indicates the robustness and reliability of the proposed code in maintaining better performance under worst-case conditions.

Moreover, the normalized average value of the PDAF (\(U_{\frac{1}{2}} (\mathbf{\Phi})\)) and the average SE (\(\overline{S}\)) for the proposed code, although not optimized, are slightly lower than those of some other codes. For instance, compared to the Frank code, our proposed code has negligible degradation in the average SE, roughly $0.225$\% for $M=16$, $1.72$\% for $M=36$, and $2.52$\% for $M=64$. This observation highlights that while the primary focus of our optimization is on enhancing the minimum values, the average metrics are not adversely affected to a large extent.

\section{Conclusion}\label{C}
In this paper, we investigate a uni-polarized RIS with a linear shape designed to transmit a common signal to multiple UEs across a wide angular region. To ensure uniform coverage, the RIS emits a broad beam with a spatially flat-like array factor, differing from the traditional narrow beam approach. We first derive the probabilistic lower and upper bounds on the average SE delivered to the UEs. Subsequently, motivated by the derived bounds, we propose a novel code for designing the phase shift of RIS elements based on maximizing the minimum PDAF across azimuth angles from \(-\frac{\pi}{2}\) to \(\frac{\pi}{2}\) using CGA to enhance the SE for the UEs. Extensive simulations evaluate the performance of our proposed code, highlighting key metrics such as the minimum and the average PDAF values and the SE delivered to the UEs.  The results show that the proposed code improves the minimum SE delivered to the UEs while maintaining the desired broad beam characteristic, outperforming established codes like the Barker, Frank, and Chu codes.

\appendices
\section{Proof of Proposition \ref{probound}}\label{app0a}
We commence with deriving the lower and upper bounds on the true expected value $E_{r,\theta}\{S\}$.
\begin{eqnarray}
\nonumber
E_{r,\theta}\{S\} &=& E_{r,\theta}\left\{\log_{2}(1 + v \beta_{g}(r) G_{0}(\theta) A(\mathbf{\Phi},\theta))\right\} \\
\nonumber
&\stackrel{(a)}{=}& E_{r}\left\{E_{\theta}\left\{\log_{2}(1 + v \beta_{g}(r) G_{0}(\theta) A(\mathbf{\Phi},\theta))\right\}\right\} \\
\nonumber
&\stackrel{(b)}{\geq}& E_{r}\left\{\log_{2}(1 + v G_{0}^{\text{min}} A^{\text{min}}(\mathbf{\Phi}) \beta_{g}(r))\right\},
\end{eqnarray}
where (a) is justified by the fact that $r$ and $\theta$ are independent, and (b) is deduced from the facts that 
\[v \beta_{g}(r) G_{0}(\theta) A(\mathbf{\Phi},\theta)>0\] 
and 
\begin{multline*}
E_{\theta}\left\{\log_{2}(1 + v \beta_{g}(r) G_{0}(\theta) A(\mathbf{\Phi},\theta))\right\} \geq \\
\log_{2}(1 + v G_{0}^{\text{min}} A^{\text{min}}(\mathbf{\Phi}) \beta_{g}(r)).   
\end{multline*}
Next, we determine the upper bound.

\begin{eqnarray}
\nonumber
E_{r,\theta}\{S\} &=& E_{r,\theta}\left\{\log_{2}(1 + v \beta_{g}(r) G_{0}(\theta) A(\mathbf{\Phi},\theta))\right\} \\
\nonumber
&\stackrel{(a)}{\leq}& \log_{2}(1 + v E_{r,\theta}\{\beta_{g}(r) G_{0}(\theta) A(\mathbf{\Phi},\theta)\}) \\
\nonumber
&\stackrel{(b)}{=}& \log_{2}(1 + v E_{r}\{\beta_{g}(r)\} E_{\theta}\{G_{0}(\theta) A(\mathbf{\Phi},\theta)\}) \\
\nonumber
&\stackrel{(c)}{\leq}& \log_{2}(1 + v G_{0}^{\text{max}} E_{r}\{\beta_{g}(r)\} E_{\theta}\{A(\mathbf{\Phi},\theta)\}),
\end{eqnarray}
where (a) leverages Jensen's inequality\footnote{Jensen's inequality states that if $f(\cdot)$ is a concave function and $X$ is a random variable, then $E\{f(X)\}\leq f(E\{X\})$.} for the concave function $\log_{2}(1 + vx)$, (b) is based on the independence of $\theta$ and $r$, and (c) uses the fact that 
\[
E_{\theta}\{ G_{0}(\theta) A(\mathbf{\Phi},\theta)\} \leq~G_{0}^{\text{max}} E_{\theta}\{A(\mathbf{\Phi},\theta)\}.
\]
Since \(r_{k}\) and \(\theta_{k}\) are uniformly distributed and independent, as previously mentioned, \(S_{1}, S_{2}, \ldots\) form a sequence of independent and identically distributed (i.i.d.) random variables with a finite mean \(E\{S\}\) and variance \(Var\{S\}\). Based on the weak law of large numbers (WLLN) \cite{SLLN}, we infer that for any $\epsilon>0$, there exists a positive integer $N_{\epsilon}$ such that when $K\geq N_{\epsilon}$, the following holds with probability $1$
\begin{equation}\label{fprob}
E_{r,\theta}\{S\}-\epsilon<\overline{S}\triangleq \frac{1}{K}\sum\limits_{k=1}^{K}S_{k}<E_{r,\theta}\{S\}+\epsilon.
\end{equation}

Ultimately, by integrating the obtained bounds with (\ref{fprob}), we ascertain the lower and upper limits for $\overline{S}$, as expressed in (\ref{uplowb}), which completes the proof of the Proposition.
\section{Proof of Lemma \ref{filem}}\label{appa}
Noting transformation (\ref{trans}), let 
\[\alpha_{1}:=\frac{2\pi \Delta}{\lambda}\left(\sin{\theta_{h}}+\sin{\theta_{1}}\right)\]
and 
\[\alpha_{2}:=\frac{2\pi \Delta}{\lambda}\left(\sin{\theta_{h}}+\sin{\theta_{2}}\right).\] 
Then, we have
\[
\left|A(\mathbf{\Phi},\theta_{2}) - A(\mathbf{\Phi},\theta_{1})\right|~~~~~~~~~~~~~~~~~~~~~~~~~~~~~~~~~~~~~~
\]
\begin{eqnarray*}
&\stackrel{(a) }{=}& \left| {\left|\sum_{m=1}^{M}e^{j\phi_{m}}e^{-j(m-1)\alpha_{2}}\right|}^{2}-{\left|\sum_{m=1}^{M}e^{j\phi_{m}}e^{-j(m-1)\alpha_{1}}\right|}^{2}\right|\\
&\stackrel{(b) }{=}& \left| \sum\limits_{n=1}^{M}\sum\limits_{m=1}^{M}e^{-j\left(\phi_{m}-\phi_{n}\right)}\left(e^{j(m-n)\alpha_{2} }-e^{j(m-n)\alpha_{1} }\right) \right |\\
&\stackrel{(c) }{\leq}& \sum\limits_{n=1}^{M}\sum\limits_{m=1}^{M}\left| e^{-j\left(\phi_{m}-\phi_{n}\right)}\right| \left| e^{j(m-n)\alpha_{2} }-e^{j(m-n)\alpha_{1} }\right|\\
&=& \sum\limits_{n=1}^{M}\sum\limits_{m=1}^{M}\sqrt{2-2\cos{\left((m-n)(\alpha_{2}-\alpha_{1})\right)}}\\
&=& \sum\limits_{n=1}^{M}\sum\limits_{m=1}^{M}2\left| \sin{\left(\frac{m-n}{2}(\alpha_{2}-\alpha_{1})\right)}\right|\\
&\stackrel{(d) }{\leq}& \sum\limits_{n=1}^{M}\sum\limits_{m=1}^{M}|m-n||\alpha_{2}-\alpha_{1}|\\
&\stackrel{(e) }{=}& \frac{(M-1)M^2\pi \Delta}{\lambda}\left|\sin{\theta_{2}}-\sin{\theta_{1}}\right|\\
&\stackrel{(f) }{\leq}& \frac{(M-1)M^2\pi \Delta}{\lambda}\left|\theta_{2}-\theta_{1} \right|,
\end{eqnarray*}
where (a) follows from (\ref{conditiono}), (b) is written from the fact that ${|Z|}^{2}=ZZ^{\star}$ for any complex value $Z$, (c) comes from the triangle inequality, (d) follows from $|\sin{x}|\leq |x|$, (e) is written from $\sum\limits_{n=1}^{M}\sum\limits_{m=1}^{M}|m-n|=\frac{(M-1)M^2}{2}$, and (f) follows from the fact that the function $\sin{(x)}$ is Lipschitz continuous with Lipschitz constant $1$.

\section{Proof of Theorem \ref{thecom}}\label{appb}
The discretization of the interval $\left[-\frac{\pi}{2}, \frac{\pi}{2}\right]$ into a finite set of points defined as $\theta_i = \frac{-\pi}{2} + \frac{\pi i}{D}$, where $i=0, 1, 2, \ldots, D$ results in a step size \( \Delta\theta = \frac{\pi}{D} \). Therefore, the distance from any point \( \theta \) within \( \left[-\frac{\pi}{2}, \frac{\pi}{2}\right] \) to the neighboring discretization points is at most \( \Delta\theta = \frac{\pi}{D} \).

Let \( \theta^\star \) denote the global minimum, and let \( \theta^c\neq \theta^\star \) represent the local minimum with the smallest value in the continuous optimization problem \( \min\limits_{\theta \in \left[-\frac{\pi}{2}, \frac{\pi}{2}\right]} A(\mathbf{\Phi},\theta) \). Since \(\theta^\star\) is the global minimum, we conclude that there exists an open interval \((\theta^\star -~\epsilon_0, \theta^\star + \epsilon_0)\) such that \(A(\mathbf{\Phi},\theta) < A(\mathbf{\Phi},\theta^c)\) for each \(\theta\) in this interval. Now, if \(\Delta \theta = \frac{\pi}{D} < 2\epsilon_0\), or equivalently \(D > \frac{\pi}{2\epsilon_0}\), at least one of the discretization points near \(\theta^\star\), say \(\theta_i^\star\), falls within \((\theta^\star - \epsilon_0, \theta^\star + \epsilon_0)\). Consequently, \(A(\mathbf{\Phi},\theta_i^\star) < A(\mathbf{\Phi},\theta^c)\), indicating that \(\theta_i^\star\) is the optimal discretization point. Therefore, we have
\begin{eqnarray*}
e&=& \left|\min_{\theta \in \left[-\frac{\pi}{2},\frac{\pi}{2}\right]} A(\mathbf{\Phi},\theta) - \min_{i=0,1,\ldots,D}A\left(\mathbf{\Phi},\frac{-\pi}{2}+\frac{\pi i}{D}\right)\right |\\
&=& \left| A(\mathbf{\Phi}, \theta^\star) - A(\mathbf{\Phi}, \theta_{i}^{\star})\right|\\
&\stackrel{(a) }{\leq} &\frac{(M-1)M^{2}\pi \Delta}{\lambda} |\theta^\star - \theta_{i}^{\star}|\\
&\stackrel{(b) }{<}& \frac{(M-1)M^{2}\pi^{2} \Delta}{\lambda D},
\end{eqnarray*}
where (a) is written from Lemma \ref{filem}, and (b) follows from the fact that the distance between $\theta^\star$ and $\theta_{i}^{\star}$ is less than the step size $\Delta \theta=\frac{\pi}{D}$. This completes the proof of the Theorem.
\section{Proof of Theorem \ref{therfin}}\label{appcc}

Noting (\ref{arrayHV}), we can write
\[
E_{\theta}\left\{A(\mathbf{\Phi},\theta)\right\}~~~~~~~~~~~~~~~~~~~~~~~~~~~~~~~~~~~~~~~~~~~~~~~~~
\]
\begin{eqnarray*}
&=&E_{\theta} \left\{{\left| \sum\limits_{m=1}^{M}e^{j\phi_{m}}e^{-j\frac{2\pi \Delta (m-1)}{\lambda}\left(\sin{\theta_{h}}+\sin{\theta}\right)}\right|}^{2}\right\}\\
&\stackrel{(a) }{=}&E_{\theta} \left\{ \sum\limits_{n=1}^{M}\sum\limits_{m=1}^{M}e^{-j\left(\phi_{m}-\phi_{n}\right)}e^{j\frac{2\pi \Delta (m-n)}{\lambda}\left(\sin{\theta_{h}}+\sin{\theta}\right)}\right\}\\
&=&\sum\limits_{n=1}^{M}\sum\limits_{m=1}^{M}e^{-j\left(\phi_{m}-\phi_{n}-a_{m-n}\sin{\theta_{h}}\right)}E_{\theta} \left\{e^{ja_{m-n} \sin{\theta} } \right\}\\
&\stackrel{(b) }{=}&\sum\limits_{n=1}^{M}\sum\limits_{m=1}^{M}J_{0}(a_{m-n})e^{-j\left(\phi_{m}-\phi_{n}-a_{m-n}\sin{\theta_{h}}\right)}\\
&=&M+\sum\limits_{n=1}^{M-1}\sum\limits_{m=n+1}^{M}J_{0}(a_{m-n})e^{-j\left(\phi_{m}-\phi_{n}-a_{m-n}\sin{\theta_{h}}\right)}\\
&~&+\sum\limits_{n=1}^{M-1}\sum\limits_{m=n+1}^{M}J_{0}(a_{m-n})e^{j\left(\phi_{m}-\phi_{n}-a_{m-n}\sin{\theta_{h}}\right)}\\
&=&2\sum_{n=1}^{M-1}\!\sum_{m=n+1}^{M}J_{0}(a_{m-n})\cos{\left(\phi_{m}-\phi_{n}-a_{m-n}\sin{\theta_{h}}\right)}\\
&~&+M,
\end{eqnarray*}
where (a) comes from the fact that ${|Z|}^{2}=ZZ^{\star}$ for any complex value $Z$, and (b) follows from (\ref{besseldef}).
\section{Proof of Proposition \ref{proppp2}}\label{appddd}
We begin by finding a closed-form expression for the $\max\limits_{\mathbf{\Phi}}E_{\theta}\left\{A(\mathbf{\Phi},\theta)\right\}$ when $\frac{\Delta}{\lambda}=\frac{1}{2}$. As $x \to +\infty$, the Bessel function of the first kind with order zero  has the asymptotic expression \cite{asymptotic}

\[J_{0}(x) = \sqrt{\frac{2}{\pi x}} \cos{\left(x - \frac{\pi}{4}\right)} + \mathcal{O}\left(\frac{1}{x^{\frac{3}{2}}}\right).\]
Therefore, we can approximate the value of $J_{0}(a_{k})$ by
\begin{equation}\label{bessapp}
J_{0}(a_{k})=J_{0}(k\pi) \approx \sqrt{\frac{2}{k{\pi}^{2}}} \cos{\left(k\pi - \frac{\pi}{4}\right)}=\frac{{(-1)}^{k}}{\pi\sqrt{k}}.
\end{equation}
We note that while this approximation may not be precise for small values of \(k\), the sign of \(J_{0}(k\pi)\) remains consistent in the approximation across all \(k\) values. That is, \(J_{0}(k\pi)\) and \(\frac{{(-1)}^{k}}{\pi\sqrt{k}}\) always share the same sign. This sign consistency is crucial for maximizing $E_{\theta}\left\{A(\mathbf{\Phi},\theta)\right\}$. Noting (\ref{bessapp}), $E_{\theta}\left\{A(\mathbf{\Phi},\theta)\right\}$ can be approximated by
\[
E_{\theta}\left\{A(\mathbf{\Phi},\theta)\right\}\approx~~~~~~~~~~~~~~~~~~~~~~~~~~~~~~~~~~~~~~~~~~~~~~~~~
\]
\[2\sum_{n=1}^{M-1}\sum_{m=n+1}^{M}\frac{{(-1)}^{m-n}}{\pi\sqrt{m-n}}\cos{\left(\phi_{m}-\phi_{n}-(m-n)\pi \sin{\theta_{h}}\right)}\]
\begin{equation*}
+M.~~~~~~~~~~~~~~~~~~~~~~~~~~~~~~~~~~~~~~~~~~~~~~~~~~~~~~~~~~~~~~~
\end{equation*}

As it can be seen, the amplitude of the cosine terms is positive when \(m-n\) is even and negative when \(m-n\) is odd. Hence, we conclude that $E_{\theta}\left\{A(\mathbf{\Phi},\theta)\right\}$ is maximized when 
\begin{eqnarray}\label{eui}
\cos{\left(\phi_{m}-\phi_{n}-(m-n)\pi \sin{\theta_{h}}\right)}&=&{(-1)}^{m-n}\\\nonumber
&=&\cos{((m-n)\pi)}.
\end{eqnarray}

It is not hard to see that the solution of  (\ref{eui}) is
\begin{equation*}
\phi_{m}^{\text{max-average}} \equiv\phi_{0}+(m-1)(1+\sin{\theta_{h}})\pi ~(\text{mod}~ 2\pi),
\end{equation*}
where $\phi_{0}\in [0,2\pi]$ is an arbitrary constant value, and $1 \leq~m \leq M$.

Finally, noting (\ref{defsef}), Theorem \ref{therfin}, and the derived maximum-average code, we can express the expanded formula for calculating $U_{\frac{1}{2}} (\mathbf{\Phi})$ as shown in (\ref{expan}).

\end{document}